\documentclass[12pt,a4paper]{article}

\usepackage{ifthen} 
\newboolean{pdflatex}
\setboolean{pdflatex}{true} 

\newboolean{articletitles}
\setboolean{articletitles}{true} 

\newboolean{uprightparticles}
\setboolean{uprightparticles}{false} 

\newboolean{inbibliography}
\setboolean{inbibliography}{false} 

\usepackage{microtype}
\usepackage{lineno}  
\usepackage{xspace} 
\usepackage{caption} 

\usepackage{graphicx}  
\usepackage{color}
\usepackage{colortbl}
\graphicspath{{./figs/}} 

\usepackage{amsmath} 
\usepackage{amssymb}
\usepackage{amsfonts}
\usepackage{upgreek} 

\newcommand*\patchAmsMathEnvironmentForLineno[1]{%
\expandafter\let\csname old#1\expandafter\endcsname\csname #1\endcsname
\expandafter\let\csname oldend#1\expandafter\endcsname\csname
end#1\endcsname
 \renewenvironment{#1}%
   {\linenomath\csname old#1\endcsname}%
   {\csname oldend#1\endcsname\endlinenomath}%
}
\newcommand*\patchBothAmsMathEnvironmentsForLineno[1]{%
  \patchAmsMathEnvironmentForLineno{#1}%
  \patchAmsMathEnvironmentForLineno{#1*}%
}
\AtBeginDocument{%
\patchBothAmsMathEnvironmentsForLineno{equation}%
\patchBothAmsMathEnvironmentsForLineno{align}%
\patchBothAmsMathEnvironmentsForLineno{flalign}%
\patchBothAmsMathEnvironmentsForLineno{alignat}%
\patchBothAmsMathEnvironmentsForLineno{gather}%
\patchBothAmsMathEnvironmentsForLineno{multline}%
\patchBothAmsMathEnvironmentsForLineno{eqnarray}%
}

\usepackage{hyperref}    
\usepackage[all]{hypcap} 

\def\MagUp {\mbox{\em Mag\kern -0.05em Up}\xspace}

\ifthenelse{\boolean{uprightparticles}}%
{

 \def\Ppi         {\ensuremath{\uppi}\xspace}

 \def\PDelta      {\ensuremath{\Delta}\xspace}                 
 \def\PXi      {\ensuremath{\Xi}\xspace}                 
 \def\PLambda      {\ensuremath{\Lambda}\xspace}                 
 \def\PSigma      {\ensuremath{\Sigma}\xspace}                 
 \def\POmega      {\ensuremath{\Omega}\xspace}                 
 \def\PUpsilon      {\ensuremath{\Upsilon}\xspace}                 
 
 \def\PB      {\ensuremath{\mathrm{B}}\xspace}                 
                  
 \def\PD      {\ensuremath{\mathrm{D}}\xspace}

 \def\PK      {\ensuremath{\mathrm{K}}\xspace}

 \def\Pb      {\ensuremath{\mathrm{b}}\xspace}                 
 \def\Pc      {\ensuremath{\mathrm{c}}\xspace}                 
                  
 \def\Pe      {\ensuremath{\mathrm{e}}\xspace}

 \def\Pi      {\ensuremath{\mathrm{i}}\xspace}

 \def\Ps      {\ensuremath{\mathrm{s}}\xspace}

}
{

 \def\Ppi         {\ensuremath{\pi}\xspace}

 \mathchardef\PDelta="7101
 \mathchardef\PXi="7104
 \mathchardef\PLambda="7103
 \mathchardef\PSigma="7106
 \mathchardef\POmega="710A
 \mathchardef\PUpsilon="7107
                  
 \def\PB      {\ensuremath{B}\xspace}                 
                  
 \def\PD      {\ensuremath{D}\xspace}

 \def\PK      {\ensuremath{K}\xspace}

 \def\Pb      {\ensuremath{b}\xspace}                 
 \def\Pc      {\ensuremath{c}\xspace}                 
                  
 \def\Pe      {\ensuremath{e}\xspace}

 \def\Pi      {\ensuremath{i}\xspace}

 \def\Ps      {\ensuremath{s}\xspace}

}

\makeatletter
\ifcase \@ptsize \relax
  \newcommand{\miniscule}{\@setfontsize\miniscule{4}{5}}
\or
  \newcommand{\miniscule}{\@setfontsize\miniscule{5}{6}}
\or
  \newcommand{\miniscule}{\@setfontsize\miniscule{5}{6}}
\fi
\makeatother

\DeclareRobustCommand{\optbar}[1]{\shortstack{{\miniscule (\rule[.5ex]{1.25em}{.18mm})}
  \\ [-.7ex] $#1$}}

\def\epem       {{\ensuremath{\Pe^+\Pe^-}}\xspace}

\def\squark    {{\ensuremath{\Ps}}\xspace}

\def\cquark    {{\ensuremath{\Pc}}\xspace}

\def\bquark    {{\ensuremath{\Pb}}\xspace}

\def\pion   {{\ensuremath{\Ppi}}\xspace}
\def\piz    {{\ensuremath{\pion^0}}\xspace}

\def\pip    {{\ensuremath{\pion^+}}\xspace}
\def\pim    {{\ensuremath{\pion^-}}\xspace}

\def\pimp   {{\ensuremath{\pion^\mp}}\xspace}

\def\kaon    {{\ensuremath{\PK}}\xspace}
  \def\Kbar    {{\kern 0.2em\overline{\kern -0.2em \PK}{}}\xspace}

\def\KorKbar    {\kern 0.18em\optbar{\kern -0.18em K}{}\xspace}
\def\Kz      {{\ensuremath{\kaon^0}}\xspace}
\def\Kzb     {{\ensuremath{\Kbar{}^0}}\xspace}
\def\Kp      {{\ensuremath{\kaon^+}}\xspace}
\def\Km      {{\ensuremath{\kaon^-}}\xspace}
\def\Kpm     {{\ensuremath{\kaon^\pm}}\xspace}
\def\Kmp     {{\ensuremath{\kaon^\mp}}\xspace}
\def\KS      {{\ensuremath{\kaon^0_{\rm\scriptscriptstyle S}}}\xspace}

\def\Kstarz  {{\ensuremath{\kaon^{*0}}}\xspace}
\def\Kstarzb {{\ensuremath{\Kbar{}^{*0}}}\xspace}
\def\Kstar   {{\ensuremath{\kaon^*}}\xspace}

  \def\Dbar    {{\kern 0.2em\overline{\kern -0.2em \PD}{}}\xspace}
\def\D       {{\ensuremath{\PD}}\xspace}

\def\DorDbar    {\kern 0.18em\optbar{\kern -0.18em D}{}\xspace}
\def\DtwoorDtwobar {\kern -0.25em\optbar{\kern 0.25em D_2^*}{}\xspace}
\def\Dz      {{\ensuremath{\D^0}}\xspace}
\def\Dzb     {{\ensuremath{\Dbar{}^0}}\xspace}

\def\Dstar   {{\ensuremath{\D^*}}\xspace}

\def\Dstarz  {{\ensuremath{\D^{*0}}}\xspace}

\def\B       {{\ensuremath{\PB}}\xspace}
\def\Bbar    {{\ensuremath{\kern 0.18em\overline{\kern -0.18em \PB}{}}}\xspace}

\def\BorBbar    {\kern 0.18em\optbar{\kern -0.18em B}{}\xspace}
\def\Bz      {{\ensuremath{\B^0}}\xspace}
\def\Bzb     {{\ensuremath{\Bbar{}^0}}\xspace}
\def\BzorBzbar  {\kern 0.18em\optbar{\kern -0.18em B}{}^0\xspace}
\def\Bu      {{\ensuremath{\B^+}}\xspace}
\def\Bub     {{\ensuremath{\B^-}}\xspace}
\def\Bp      {{\ensuremath{\Bu}}\xspace}
\def\Bm      {{\ensuremath{\Bub}}\xspace}

\def\Bs      {{\ensuremath{\B^0_\squark}}\xspace}
\def\Bsb     {{\ensuremath{\Bbar{}^0_\squark}}\xspace}

  \def\Y#1S{\ensuremath{\PUpsilon{(#1S)}}\xspace}

\def\FourS {{\Y4S}}

\def\Lbar        {{\ensuremath{\kern 0.1em\overline{\kern -0.1em\PLambda}}}\xspace}
\def\LorLbar    {\kern 0.18em\optbar{\kern -0.18em \PLambda}{}\xspace}

\def\to                 {\ensuremath{\rightarrow}\xspace}

\def\CP                {{\ensuremath{C\!P}}\xspace}

\def\AT#1     {\ensuremath{A_{\mathrm{T}}^{#1}}\xspace}           

\def\C#1      {\ensuremath{\mathcal{C}_{#1}}\xspace}                       
\def\Cp#1     {\ensuremath{\mathcal{C}_{#1}^{'}}\xspace}                    
\def\Ceff#1   {\ensuremath{\mathcal{C}_{#1}^{\mathrm{(eff)}}}\xspace}        
\def\Cpeff#1  {\ensuremath{\mathcal{C}_{#1}^{'\mathrm{(eff)}}}\xspace}       
\def\Ope#1    {\ensuremath{\mathcal{O}_{#1}}\xspace}                       
\def\Opep#1   {\ensuremath{\mathcal{O}_{#1}^{'}}\xspace}                    

\newcommand{\tev}{\ifthenelse{\boolean{inbibliography}}{\ensuremath{~T\kern -0.05em eV}\xspace}{\ensuremath{\mathrm{\,Te\kern -0.1em V}}}\xspace}
\newcommand{\gev}{\ensuremath{\mathrm{\,Ge\kern -0.1em V}}\xspace}
\newcommand{\mev}{\ensuremath{\mathrm{\,Me\kern -0.1em V}}\xspace}
\newcommand{\kev}{\ensuremath{\mathrm{\,ke\kern -0.1em V}}\xspace}
\newcommand{\ev}{\ensuremath{\mathrm{\,e\kern -0.1em V}}\xspace}
\newcommand{\gevc}{\ensuremath{{\mathrm{\,Ge\kern -0.1em V\!/}c}}\xspace}
\newcommand{\mevc}{\ensuremath{{\mathrm{\,Me\kern -0.1em V\!/}c}}\xspace}
\newcommand{\gevcc}{\ensuremath{{\mathrm{\,Ge\kern -0.1em V\!/}c^2}}\xspace}
\newcommand{\gevgevcccc}{\ensuremath{{\mathrm{\,Ge\kern -0.1em V^2\!/}c^4}}\xspace}
\newcommand{\mevcc}{\ensuremath{{\mathrm{\,Me\kern -0.1em V\!/}c^2}}\xspace}

\def\invfb   {\ensuremath{\mbox{\,fb}^{-1}}\xspace}

\def\invab   {\ensuremath{\mbox{\,ab}^{-1}}\xspace}

\def\gsim{{~\raise.15em\hbox{$>$}\kern-.85em
          \lower.35em\hbox{$\sim$}~}\xspace}
\def\lsim{{~\raise.15em\hbox{$<$}\kern-.85em
          \lower.35em\hbox{$\sim$}~}\xspace}

\def\tell1  {TELL1\xspace}
\def\ukl1   {UKL1\xspace}

\newcommand{\eg}{\mbox{\itshape e.g.}\xspace}
\newcommand{\ie}{\mbox{\itshape i.e.}\xspace}

\usepackage{cite} 
\usepackage{mciteplus}

\usepackage{rotating} 
\usepackage{arydshln} 
\usepackage{relsize} 
\usepackage[hang,flushmargin]{footmisc}

\begin{document}

\renewcommand{\thefootnote}{\fnsymbol{footnote}}
\setcounter{footnote}{1}

\begin{titlepage}
\pagenumbering{roman}

{\bf\boldmath\huge
\begin{center}
  Optimising sensitivity to $\gamma$ with $\Bz \to D\Kp\pim$, $D \to \KS\pip\pim$ double Dalitz plot analysis
\end{center}
}

\vspace*{2.0cm}

\begin{center}
  D.~Craik$^1$, T.~Gershon$^2$, A.~Poluektov$^2$
\bigskip\\
{\it\footnotesize 
$ ^1$ Massachusetts Institute of Technology, Cambridge, MA, United States\\
$ ^2$ Department of Physics, University of Warwick, Coventry, United Kingdom\\
}
\end{center}

\vspace{\fill}

\begin{abstract}
  \noindent
  Two of the most powerful methods currently used to determine the angle $\gamma$ of the CKM Unitarity Triangle exploit $\Bp \to D\Kp$, $D \to \KS\pip\pim$ decays and $\Bz \to D\Kp\pim$, $D \to \Kp\Km$, $\pip\pim$ decays.
  It is possible to combine the strengths of both approaches in a ``double Dalitz plot'' analysis of $\Bz \to D\Kp\pim$, $D \to \KS\pip\pim$ decays.
  The potential sensitivity of such an analysis is investigated in the light of recently published experimental information on the $\Bz \to D\Kp\pim$ decay.
  The formalism is also expanded, compared to previous discussions in the literature, to allow $\Bz \to D\Kp\pim$ with any subsequent $D$ decay to be included. 
\end{abstract}

\vspace{\fill}

\end{titlepage}

\newpage
\setcounter{page}{2}
\mbox{~}

\cleardoublepage

\renewcommand{\thefootnote}{\arabic{footnote}}
\setcounter{footnote}{0}

\pagestyle{plain} 
\setcounter{page}{1}
\pagenumbering{arabic}

\section{Introduction}
\label{sec:intro}

Within the Standard Model, the sole source of \CP\ violation is the complex phase of the Cabibbo-Kobayashi-Maskawa (CKM) quark mixing matrix~\cite{Cabibbo:1963yz,Kobayashi:1973fv}.
The amount of matter-antimatter asymmetry related to this source can be quantified through the area of the Unitarity Triangle formed from elements of the CKM quark mixing matrix~\cite{Jarlskog:1985ht}.
The angle $\gamma \equiv \arg\left[-V_{ud}^{}V_{ub}^*/(V_{cd}^{}V_{cb}^*)\right]$ of this triangle is a particularly important parameter, since it can be determined with negligible theoretical uncertainty~\cite{Brod:2013sga} using methods that are reliable in the Standard Model and in any extensions that do not affect tree-level $b$ hadron decays\cite{Brod:2014bfa}. 
The current world average value is $\gamma = (76.2 \,^{+4.7}_{-5.0})^\circ$~\cite{HFLAV}, dominated by recent results from LHCb~\cite{LHCb-PAPER-2016-003,LHCb-PAPER-2014-041,LHCb-PAPER-2017-021,LHCb-PAPER-2016-032,LHCb-CONF-2017-004}.
The uncertainty is still far from the sub-degree precision that is strived for, and therefore improving the measurement of $\gamma$ remains one of the main objectives of current and planned flavour physics experiments~\cite{LHCb-PAPER-2012-031,CERN-LHCC-2017-003,Aushev:2010bq}.

Numerous variations of methods to determine $\gamma$ have been proposed, and a significant number have now been attempted experimentally (see reviews in Refs.~\cite{HFLAV,Gershon:2016fda}).
In this work, the focus is on methods based on Dalitz plot analysis of $\Bz \to D\Kp\pim$ decays~\cite{Gershon:2008pe,Gershon:2009qc}, where the neutral \D\ meson is reconstructed in final states to which both \Dz\ and \Dzb\ can decay.\footnote{
  The symbol $\D$ is used to refer to a neutral charm meson that is any admixture of $\Dz$ and $\Dzb$ states.
}
The Dalitz plot contains resonant and nonresonant contributions, including those for $\Bz \to D\Kstar(892)^0$ and $\Bz \to \D_2^*(2460)^-\Kp$ decays.
In the $\Bz \to D\Kstarz$ case,\footnote{
  Throughout this paper the symbol $\Kstar$ will be used to denote the $\Kstar(892)$ resonance unless explicitly stated otherwise.}
the amplitudes from $b \to c$ transitions can interfere with those from $b \to u$ transitions, and \CP-violating observables are related to their relative weak (\ie, \CP-violating) and strong (\ie, \CP-conserving) phases $\gamma$ and $\delta_B$, as well as their relative magnitude, $r_B$.
One advantage of using neutral $B$ meson decays to determine $\gamma$, compared to the more familiar approach with $\Bp \to D\Kp$ decays~\cite{Gronau:1990ra,Gronau:1991dp,Atwood:1996ci,Atwood:2000ck}, is that the value of $r_B$ associated with $\Bz \to D\Kstarz$ transitions is expected to be larger (typical expectations are $r_B(D\Kstarz) \sim 0.3$, $r_B(D\Kp) \sim 0.1$, while the latest world averages are $r_B(D\Kstarz) = 0.226 \,^{+0.042}_{-0.045}$, $r_B(D\Kp) \sim 0.105 \pm 0.005$~\cite{HFLAV}).
Another advantage of the Dalitz plot analysis approach is  that interference effects between the amplitudes for $D\Kstarz$ and contributions such as $\D_2^*(2460)^-\Kp$ involving $D\pim$ resonances, which are mediated by $b \to c$ transitions only, can be used to enhance the sensitivity and resolve ambiguities in the allowed values of $\gamma$~\cite{Gershon:2008pe}.

The LHCb collaboration has recently performed the first determination of $\gamma$ with $\Bz \to D\Kp\pim$ Dalitz plot analysis, using \D\ meson decays to $\Kp\Km$ and $\pip\pim$~\cite{LHCb-PAPER-2015-059}, building on knowledge of the $\Bz \to \Dzb\Kp\pim$ Dalitz plot structure obtained in an earlier analysis (with $\Dzb \to \Kp\pim$)~\cite{LHCb-PAPER-2015-017}.
The precision obtained on the parameters $x_\pm = r_B \cos(\delta_B \pm \gamma)$ and $y_\pm = r_B \sin(\delta_B \pm \gamma)$ is comparable~\cite{Gershon:2017kqy} to that from analysis of $\Bz \to D\Kstarz$, $D \to \KS\pip\pim$ decays~\cite{LHCb-PAPER-2016-006,LHCb-PAPER-2016-007} selected from the same data sample.
This demonstrates the potential impact of the $\Bz \to D\Kp\pim$ Dalitz plot technique on the determination of $\gamma$.
In the latter analysis a ``quasi-two-body'' approach is used, in which the $\Kstarz$ resonance is treated as a stable particle and the effects of other contributions in the selected region of the $D\Kp\pim$ Dalitz plot are absorbed in hadronic parameters~\cite{Gronau:2002mu}.
A further advantage of the Dalitz plot analysis is that this treatment is not necessary, and moreover the extra hadronic parameters that enter in the quasi-two-body approach can be measured. 

The results of the Dalitz plot analysis~\cite{LHCb-PAPER-2015-059} however suffer from two important sources of systematic uncertainty.  
The first is that the modelling of the suppressed and favoured amplitudes in $\Bz \to D\Kp\pim$ decays impacts the obtained results.
While narrow resonances such as the $\Kstar(892)^0$ and $\D_2^*(2460)^-$ states can be reliably described by relativistic Breit--Wigner functions, there are also broad (\eg\ $\kaon^*_0(1430)^0$ and $\D_0^*(2400)^-$) and possible nonresonant contributions for which an appropriate range of alternative lineshapes must be considered.
The second is due to background from $\Bsb \to \Dstar\Kp\pim$, where the soft pion or photon from $\Dstar \to D\piz$ or $D \gamma$ is not included in the reconstruction, which peaks near to the signal region. 
Since the favoured final state for the \Bz\ decay is suppressed for the \Bs\ decay, and vice versa, this particularly impacts the $\Bz \to D\Kp\pim$, $D \to \Km\pip$ channel (which, for this reason, was not included in the LHCb analysis~\cite{LHCb-PAPER-2015-059}), but is also important for the $\Bz \to D\Kp\pim$, $D \to \Kp\Km$ and $\pip\pim$ modes.
It should be noted, however, that this issue would not affect analyses performed on data samples collected using the $\epem \to \FourS \to B\bar{B}$ process, such as those that will be available in the Belle~II experiment, since there is no production of $\Bs$ mesons in that case.

Both of these effects suggest that a promising way to proceed may be via model-independent double Dalitz plot analysis of the $\Bz \to D\Kp\pim$, $D \to \KS\pip\pim$ decay.
This method, introduced in Ref.~\cite{Gershon:2009qr}, builds on ideas introduced for $\Bp \to D\Kp$, $D \to \KS\pip\pim$ decays~\cite{Giri:2003ty,Bondar,Bondar:2005ki,Bondar:2008hh}, where the \D\ decay Dalitz plot is divided into bins.
Each of the bins is described by hadronic parameters corresponding approximately to the average cosine or sine of the strong phase difference between the amplitudes for \Dz\ and \Dzb\ decays in that bin.
Together with external input on the \D\ decay hadronic parameters, as can be (and has been) obtained from $\psi(3770) \to \Dz\Dzb$ data~\cite{Libby:2010nu}, only the yield in each bin needs to be determined from \B\ decay data in order to have sensitivity to $\gamma$.
The key additional ingredient in the double Dalitz plot analysis is that the \B\ decay Dalitz plot can also be binned, and that the corresponding \B\ decay hadronic parameters can be determined from the data simultaneously with $\gamma$ with no additional external information required.  
Additional \D\ decays, such as $D \to \Kp\Km$ and $\pip\pim$ can be included in the analysis and provide extra sensitivity, but the three-body \D\ decay is necessary in order for the method to work.
Thus, in this paper the phrase ``model-independent double Dalitz plot analysis'' refers to the study of $\Bz \to D\Kp\pim$ decays with any set of \D\ meson decays that includes $D \to \KS\pip\pim$.

The model-independent double Dalitz plot analysis approach not only resolves the issue of model-dependency, but also ameliorates the challenges presented by the $\Bsb \to \Dstar\Kp\pim$ background because a detailed description of the phase-space distribution of this decay is no longer required.
Instead, only the shape of the background in the $D\Kp\pim$ invariant mass need be described.
Recent results from LHCb have demonstrated how this can be achieved~\cite{LHCb-PAPER-2017-021}.

Consequently, it is timely to re-examine the potential of the model-independent double Dalitz plot analysis to determine $\gamma$.
This allows the study of Ref.~\cite{Gershon:2009qr} to be updated, incorporating information about $\Bz \to D\Kp\pim$ decays that is now available, and also with more realistic estimates of the yields that should be available at LHCb after the completion of LHC Run~II, and with larger data samples.
In addition, the previous study considered as a baseline including only the $D \to \KS\pip\pim$ mode together with the favoured $D \to \Kp\pim$ channel for normalisation, with the impact of adding $D \to \Kp\Km$ and $\pip\pim$ decays also assessed.  
The updated study presented here also considers inclusion of the suppressed $D \to \Km\pip$ decay.
Indeed, the formalism set out in Sec.~\ref{sec:formalism} allows any $D$ decay mode to be included in the analysis.
An estimate of the potential sensitivity, and its dependence on sample size, binning of the Dalitz plot, inclusion of different $D$ decay modes and on the impact of the $\Bsb \to \Dstar\Kp\pim$ background is presented in Sec.~\ref{sec:toy}.
A summary concludes the paper in Sec.~\ref{sec:summary}.

\section{Formalism}
\label{sec:formalism}

Following Ref.~\cite{Gershon:2009qr}, it is useful to begin by recalling the essentials of the $\Bp \to D\Kp$, $D \to \KS\pip\pim$ model-independent method~\cite{Giri:2003ty,Bondar,Bondar:2005ki,Bondar:2008hh}.
The amplitude of the decay is written as a function of \D\ decay Dalitz plot co-ordinates $(m_+^2,m_-^2)\equiv(m^2_{\KS\pi^+}, m^2_{\KS\pi^-})$,
\begin{equation}
  A_{D\,{\rm Dlz}} = \overline{A}_D + r_B e^{i(\delta_B+\gamma)} A_D\,, 
  \label{eq:AD-Dlz}
\end{equation}
where $\overline{A}_D =\overline{A}_D(m_+^2,m_-^2)$ is the amplitude of the $\Dzb \to \KS\pip\pim$ decay, and $A_D=A_D(m_+^2,m_-^2)$ is the amplitude of the $\Dz \to \KS\pip\pim$ decay.
(The favoured $B$ decay amplitude, which multiplies the right-hand side of Eq.~(\ref{eq:AD-Dlz}), is conventionally omitted as it does not affect the observables of interest.)
Assuming no \CP\ violation in $D$ decay, $A_D(m_+^2,m_-^2) = \overline{A}_D(m^2_-,m^2_+)$.\footnote{
  Effects due to \CP\ violation in $D$ decay are known to be sufficiently small that they can be neglected~\cite{HFLAV}.
  Charm mixing effects are more important, but it is well-known how to take them into account in the analysis~\cite{Silva:1999bd,Rama:2013voa,LHCb-PAPER-2013-020,LHCb-PAPER-2016-032}, thus they are not considered in this paper.
  Effects due to \CP\ violation in the $\Kz$--$\Kzb$ system are also negligible~\cite{Grossman:2013woa}.
}
The density of the $D$ decay Dalitz plot from $\Bp \to D\Kp$ decay is then given by 
\begin{equation}
  |A_{D\,{\rm Dlz}}|^2  = 
  \left|\overline{A}_D\right|^2+r_B^2\left|A_D\right|^2+2\left|A_D\right|\left|\overline{A}_D\right|(x_+c-y_+s)\,,
\end{equation}
where the functions $c=c(m_+^2,m_-^2)$ and $s=s(m_+^2,m_-^2)$ are the cosine and sine of the strong phase difference $\delta_D(m_+^2,m_-^2)=\arg A_D(m_+^2,m_-^2)-\arg \overline{A}_D(m_+^2,m_-^2)$ between the $\Dz \to \KS\pip\pim$ and $\Dzb \to \KS\pip\pim$ amplitudes.
The parameters $x_\pm = r_B \cos(\delta_B \pm \gamma)$ and $y_\pm = r_B \sin(\delta_B \pm \gamma)$ are those defined in Sec.~\ref{sec:intro}.
The equations for the charge-conjugate mode $\Bm \to D \Km$ are obtained with the substitution $\gamma \longrightarrow -\gamma$, \ie\ $(x_+,y_+) \longrightarrow (x_-,y_-)$, and $\overline{A}_D \longleftrightarrow A_D$. 
Considering both $B$ charges, one can obtain $\gamma$ and $\delta_B$ separately.

Once the Dalitz plot is divided into $2\mathcal{N}$ bins symmetrically to the exchange $m^2_-\leftrightarrow m^2_+$, the expected number of events in the $i^{\rm th}$ bin of the $D \to \KS\pip\pim$ Dalitz plot from $\Bp \to D\Kp$ decay is
\begin{equation}
  \langle N_i\rangle = 
  h_{D\,{\rm Dlz}} \left[
    K_i + r_B^2K_{-i} + 2\sqrt{K_iK_{-i}}(x_+c_i-y_+s_i)
  \right] \,, 
  \label{eq:n_b}
\end{equation}
where $h_{D\,{\rm Dlz}}$ is a normalisation constant.
The bin index $i$ ranges from $-\mathcal{N}$ to $\mathcal{N}$ (excluding 0); the exchange $m^2_+ \leftrightarrow m^2_-$ corresponds to the exchange $i\leftrightarrow -i$. 
The per-bin coefficients $c_i$ and $s_i$ are given by
\begin{equation}
  c_i = \frac{
    \int\limits_{\mathcal{D}_i} \left|A_D\right|\left|\overline{A}_D\right| \cos\delta_D\,d\mathcal{D}
  }{
    \sqrt{
      \int\limits_{\mathcal{D}_i}\left|A_D\right|^2 d\mathcal{D}
      \int\limits_{\mathcal{D}_i}\left|\overline{A}_D\right|^2 d\mathcal{D}
    }}\,, 
  \qquad \qquad
  s_i = \frac{
    \int\limits_{\mathcal{D}_i} \left|A_D\right|\left|\overline{A}_D\right| \sin\delta_D\,d\mathcal{D}
  }{
    \sqrt{
      \int\limits_{\mathcal{D}_i}\left|A_D\right|^2 d\mathcal{D}
      \int\limits_{\mathcal{D}_i}\left|\overline{A}_D\right|^2 d\mathcal{D}
    }}\,.
  \label{eq:cs}
\end{equation}
Here $\mathcal{D}$ represents the $D \to \KS\pip\pim$ Dalitz plot phase space and $\mathcal{D}_i$ is the bin region over which the integration is performed. 
The definitions of Eq.~(\ref{eq:cs}) imply the presence of a physical boundary, $c_i^2 + s_i^2 \leq 1$.

Equation~(\ref{eq:n_b}) also contains per-bin coefficients $K_i$, which can be obtained from the numbers of events in the corresponding bins of the Dalitz plot where the $D$ meson is in a flavour eigenstate.
Experimentally these can be obtained using $D^{*\pm}\to D\pi^\pm$ samples, where the charge of the emitted pion in the $\Dstar$ decay tags the flavour of the $D$ meson.

For $\Bz \to D\Kp\pim$ decays, the variation of the amplitudes $\overline{A}_B$ for $\Bz \to \Dzb\Kp\pim$ decay and $A_B$ for $\Bz \to \Dz\Kp\pim$ decay across the phase-space described by $(m^2_{D\pi},m^2_{K\pi})$ must be considered.
For simplicity, the relative weak phase $\gamma$ is factored out in the expressions that follow.  
The replacement for Eq.~(\ref{eq:AD-Dlz}) is then 
\begin{equation}
  A_{{\rm dbl}\,{\rm Dlz}} = 
  \overline{A}_B\overline{A}_D + e^{i\gamma} A_B A_D\,, 
  \label{eq:Adbl-Dlz}
\end{equation}
giving 
\begin{eqnarray}
  |A_{{\rm dbl}\,{\rm Dlz}}|^2  & = &  
  \left|\overline{A}_B\right|^2\left|\overline{A}_D\right|^2 + \left|A_B\right|^2\left|A_D\right|^2 \\
  &&
  {} + 2\left|\overline{A}_B\right|\left|\overline{A}_D\right|\left|A_B\right|\left|A_D\right|
  \left[
    (\varkappa c - \sigma s)\cos\gamma - (\varkappa s + \sigma c)\sin\gamma 
  \right] \, , \nonumber
\end{eqnarray}
where $\varkappa$ and $\sigma$ are the cosine and sine of $\delta_B = \arg(A_B) - \arg(\overline{A}_B)$, and are functions of $B$ decay Dalitz plot position.
Then, after integrating over the phase-space of both the $B$ and $D$ decay Dalitz plot bins (with the former denoted by the index $\alpha$, $1\le \alpha\le \mathcal{M}$, and the latter by roman indices as before), the number of expected events in each bin is
\begin{eqnarray}
  \langle N_{\alpha i} \rangle & = & h_{{\rm dbl}\,{\rm Dlz}} \Big\{ \overline{\kappa}_{\alpha}K_{i} + \kappa_{\alpha}K_{-i} \label{eq:num_bin} \\
  &&
  {} + 2 \sqrt{\kappa_{\alpha}K_{i}\overline{\kappa}_{\alpha}K_{-i}} \left[
    (\varkappa_{\alpha}c_i - \sigma_{\alpha}s_i)\cos\gamma - 
    (\varkappa_{\alpha}s_i + \sigma_{\alpha}c_i)\sin\gamma 
  \right] \Big\} \, , \nonumber
\end{eqnarray}
where the $B$ Dalitz plot bin phase terms are defined as 
\begin{equation}
  \varkappa_{\alpha} = \frac{
    \int\limits_{\mathcal{D}_{\alpha}} |A_B||\overline{A}_B|\cos\delta_B\,d\mathcal{D}
  }{
    \sqrt{
      \int\limits_{\mathcal{D}_{\alpha}}|A_B|^2 d\mathcal{D}
      \int\limits_{\mathcal{D}_{\alpha}}|\overline{A}_B|^2 d\mathcal{D}
    }}\,, 
  \qquad \qquad
  \sigma_{\alpha} = \frac{
    \int\limits_{\mathcal{D}_{\alpha}} |A_B||\overline{A}_B|\sin\delta_B\,d\mathcal{D}
  }{
    \sqrt{
      \int\limits_{\mathcal{D}_{\alpha}}|A_B|^2 d\mathcal{D}
      \int\limits_{\mathcal{D}_{\alpha}}|\overline{A}_B|^2 d\mathcal{D}
    }}\,.
  \label{kappasigma}
\end{equation}
The true values of $\varkappa_{\alpha}$ and $\sigma_{\alpha}$ must satistfy $\varkappa_{\alpha}^2 + \sigma_{\alpha}^2 \leq 1$.
The corresponding expression to Eq.~(\ref{eq:num_bin}) for $\Bzb \to D\Km\pip$ decays is obtained with the substitution $\gamma \longrightarrow -\gamma$.

The term $h_{{\rm dbl}\,{\rm Dlz}}$ that appears in Eq.~(\ref{eq:num_bin}) is a normalisation constant.
The factors $\overline{\kappa}_{\alpha}$ and $\kappa_{\alpha}$ are the $B$ decay equivalents of the $K_i$ factors for the $D$ decay, but in this case there is no convenient independent control sample from which they can be obtained.
Consequently, they must be determined as part of the analysis.
Another important difference between the $B$ and $D$ Dalitz plots is that there is no symmetry inherent in the $B$ decay since it is does not have a self-conjugate final state.
This is reflected by the bin indices running from $1\le \alpha\le \mathcal{M}$ for the $B$ decay, in contrast to the choice $-\mathcal{N} \le i \le \mathcal{N}$ (excluding zero) for the $D$ decay.

With $\mathcal{M}$ bins in the $\Bz \to D\Kp\pim$ Dalitz plot and $2\mathcal{N}$ bins in the $D \to \KS\pip\pim$ Dalitz plot, then the number of equations represented by Eq.~(\ref{eq:num_bin}) and the charge-conjugate equivalent is $4\mathcal{MN}$.
For each of the $\mathcal{M}$ $B$ decay Dalitz plot bins there are four unknown quantities to be determined: $\overline{\kappa}_{\alpha}$, $\kappa_{\alpha}$, $\varkappa_{\alpha}$ and $\sigma_{\alpha}$.
Similarly, for each of the $\mathcal{N}$ $D$ decay Dalitz plot bins there are factors of $K_{i}$, $K_{-i}$, $c_i$ and $s_i$ (after using $c_i = c_{-i}$ and $s_i = -s_{-i}$); however $K_{i}$, $K_{-i}$ can be precisely determined from independent samples and $c_i$ and $s_i$ have been measured from $\psi(3770) \to \Dz\Dzb$ data~\cite{Libby:2010nu}.
Consequently, these should not be considered ``unknown quantities'', but can be allowed to vary within their uncertainties in the analysis.  
Finally, there are two global unknown parameters: the normalisation factor $h_{{\rm dbl}\,{\rm Dlz}}$ and $\gamma$.
Thus, not counting the parameters associated with $D$ decays, there are in total $4\mathcal{M}+2$ quantities to be determined from the data.\footnote{
  The discussion here differs from that in Ref.~\cite{Gershon:2009qr}, where $\overline{\kappa}_{\alpha}$ and $\kappa_{\alpha}$ were considered to be independently known through the favoured $D \to \Kp\pim$ decay mode, and $c_i$ and $s_i$ were considered to be unknown.
}
Since typically $\mathcal{N} = 8$ is used for $D \to \KS\pip\pim$ decays, the
system can in principle be solved for any value of $\mathcal{M}$.

The discussion above has been in the context of $D \to \KS\pip\pim$ decays, but is in fact valid, with appropriate choices of the hadronic parameters, for any $D$ decay.
Thus, for example, the $D \to \KS\Kp\Km$ channel can be trivially included in the analysis: its inclusion is equivalent to simply adding more bins corresponding to different regions of $D$ decay phase space, though in practice it will also be convenient to allow different normalisation factors $h$ for the different $D$ decay modes.
Inclusion of the two-body decays $D \to \Kp\Km$ and $\pip\pim$ corresponds to a single bin with $c_i = 1, s_i = 0$ and $K_i = K_{-i}$ (in this case, the $K$ factors can be conveniently absorbed into the normalisation).
For a \CP-odd eigenstate one would have $c_i = -1$, $s_i = 0$, while for so-called quasi-\CP-eigenstates $c_i$ takes the value of the net \CP content (discussion of quasi-\CP-eigenstates can be found, for example, in Refs.~\cite{Nayak:2014tea,LHCb-PAPER-2015-014,Gershon:2015xra}).
The suppressed $D \to \Km\pip$ decay can be included with $K_i/K_{-i} = r_{K\pi}^2$, $c_i = \cos \delta_{K\pi}$ and $s_i = \sin \delta_{K\pi}$, while for the favoured $D \to \Kp\pim$ decay one should have instead $K_i/K_{-i} = r_{K\pi}^{-2}$, $c_i = \cos \delta_{K\pi}$ and $s_i = -\sin \delta_{K\pi}$.
Here, $r_{K\pi} = 0.0590 \pm 0.0003$ and $\delta_{K\pi} = \left( 15 \, ^{+8}_{-10} \right)^\circ$~\cite{HFLAV}, are the relative magnitude and phase of the suppressed and favoured $D$ decay amplitudes to the $\Kpm\pimp$ final states (note that care needs to be taken to ensure consistent phase conventions).
Similar expressions can be used for multibody suppressed/favoured pairs of modes such as $\D \to \Kpm\pimp\piz$ with the coherence factor included in the relations for $c_i$ and $s_i$~\cite{Atwood:2000ck,Lowery:2009id,LHCb-PAPER-2015-014,Gershon:2015xra}.
Relevant expressions for any other $D$ decay modes can easily be obtained.

\section{Sensitivity study}
\label{sec:toy}

\begin{figure}[htb]
  \centering
  \includegraphics[width=0.45\textwidth]{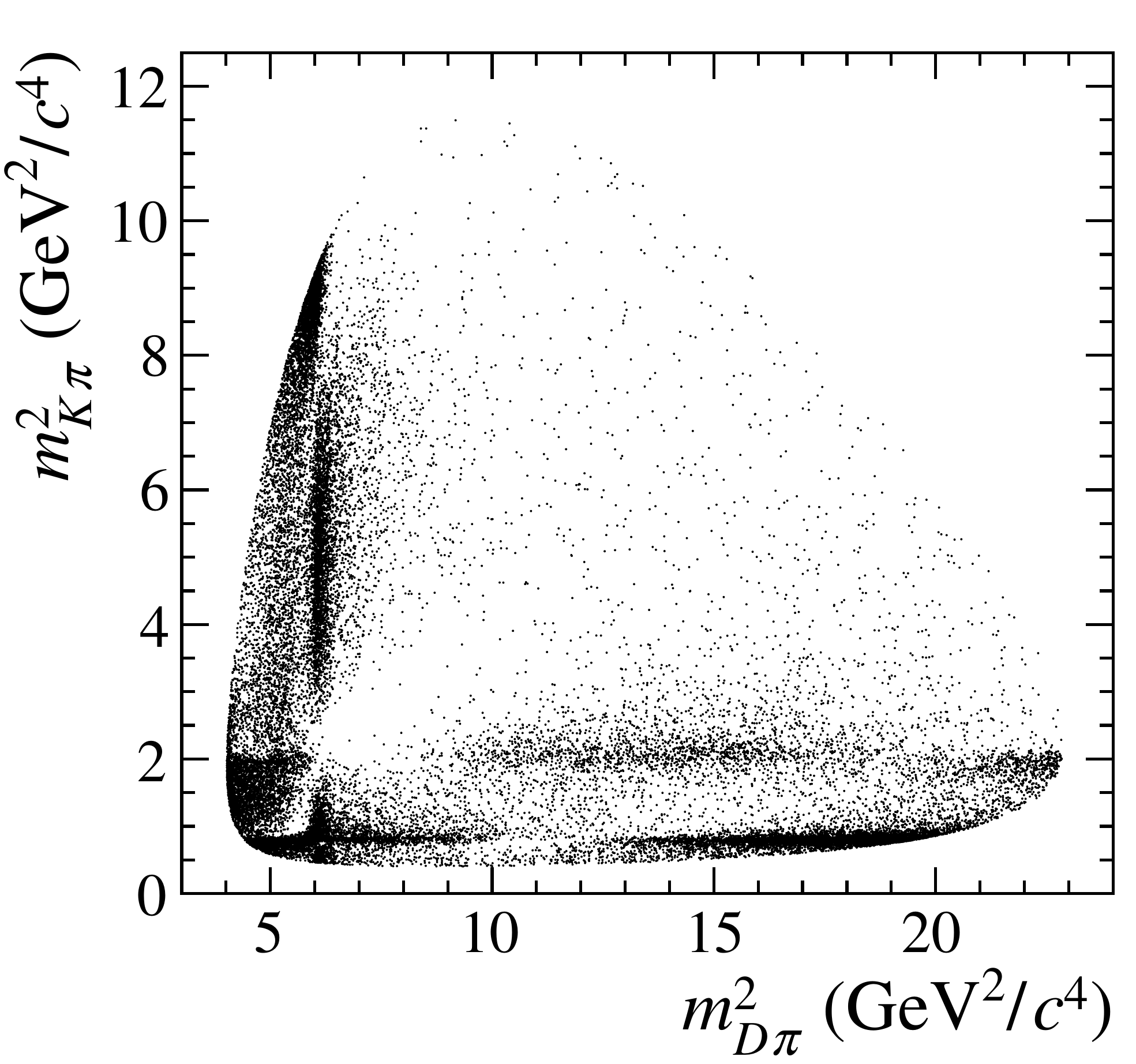}
  \put(-35, 155){(a)}
  \hspace{0.05\textwidth}
  \includegraphics[width=0.45\textwidth]{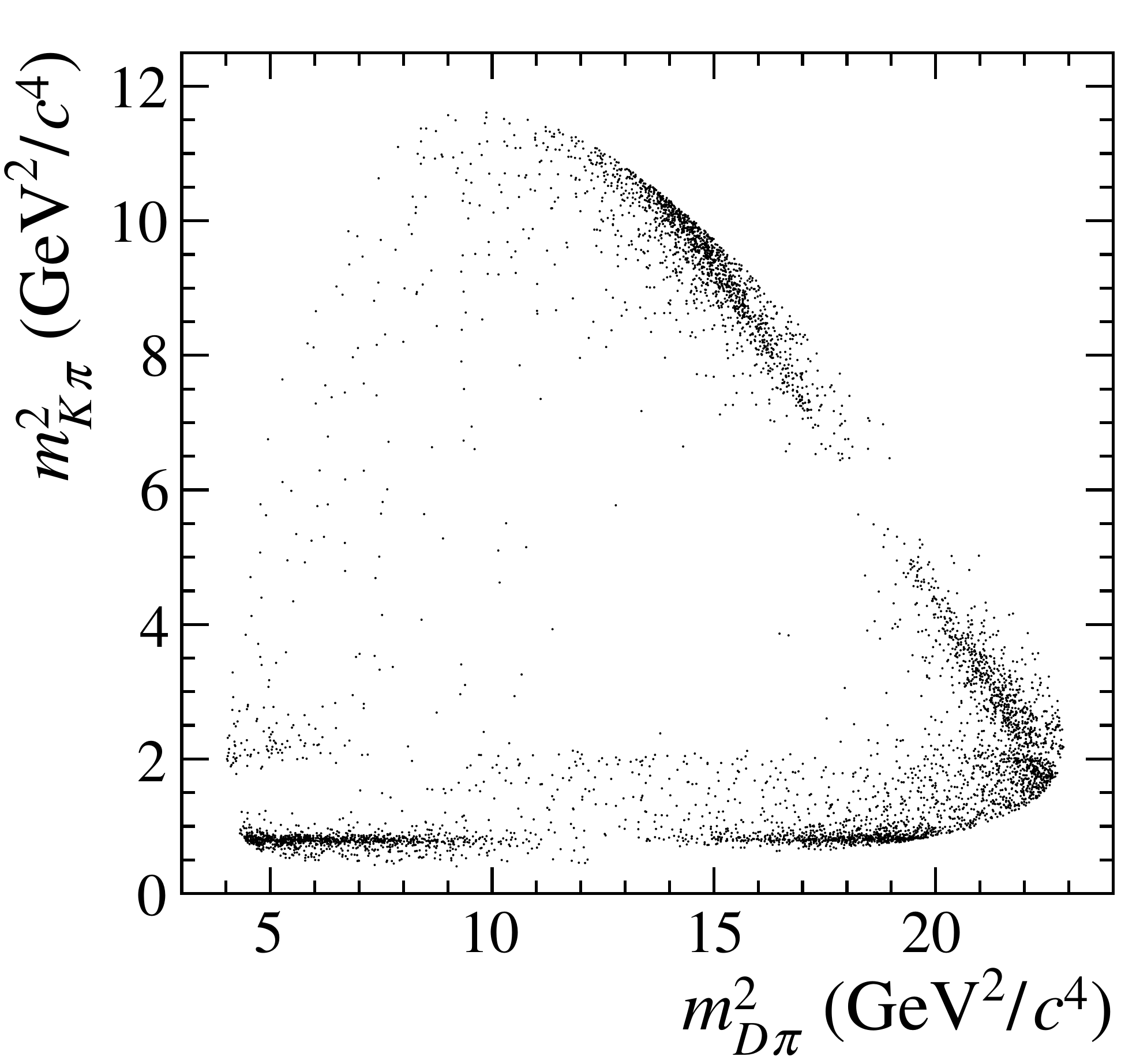}
  \put(-35, 155){(b)}
  \caption{
    Dalitz plot distributions obtained from the models used for the $B$ decay amplitudes for the (left) favoured amplitude, \ie\ $\Bz \to \Dzb \Kp\pim$, and (right) suppressed amplitude, \ie\ $\Bz \to \Dz\Kp\pim$.
    The former (latter) contains $D\pim$ ($D\Kp$) resonances and does not contain $D\Kp$ ($D\pim$) structures; both contain $\Kp\pim$ resonances.
  }
  \label{fig:toy1}
\end{figure}

In order to estimate the sensitivity of the method, simulated pseudoexperiments are generated and fitted.
To generate the Dalitz plot distributions, the favoured $B$ decay amplitude corresponds to that in the LHCb publications~\cite{LHCb-PAPER-2015-017,LHCb-PAPER-2015-059}, and the $D$ Dalitz plot model is that from Ref.~\cite{Poluektov:2010wz}.
In the baseline model, the suppressed $B$ decay amplitude is generated fixing the ratio of magnitudes of suppressed and favoured amplitudes $r_B$ to 0.3 for all of the $K^*(892)^0$, $K^*(1410)^0$, $K_2^*(1430)^0$ and $K\pi$ S-wave contributions, while taking the relative phases from the LHCb results~\cite{LHCb-PAPER-2015-059}.
The suppressed $B$ decay amplitude also includes a $D_{s1}^*(2700)^+$ component at the level indicated by the results of Ref.~\cite{LHCb-PAPER-2015-059}.
Dalitz plot distributions obtained by generating with only the favoured or suppressed $B$ decay amplitude are shown in Fig.~\ref{fig:toy1}.

Samples sizes generated correspond roughly to the expected yields at LHCb after the completion of Run~II and after accumulating $50 \invfb$ of $pp$ collision data at the end of the upgrade; these are given in Table~\ref{tab:samples}.
The expected $\Bz \to D\Kp\pim$ yields in the $D \to \Kp\pim$, $\Kp\Km$ and $\pip\pim$ channels are extrapolated from those obtained in Run~I~\cite{LHCb-PAPER-2015-059}.\footnote{
  Ref.~\cite{LHCb-PAPER-2015-059} presents yields in the signal region in bins with varying background levels.
  The background-dominated bin has been excluded from the yields presented in Table~\ref{tab:samples}.
}
The expected yield in the $D \to \Km\pip$ channel is obtained from the model assuming the same experimental efficiency, and hence normalisation factor, as for the $D \to \Kp\pim$ mode.
In the $D \to \KS\pip\pim$ channel, the expected yields also include an extrapolation from the published Run~I yields for $\Bz \to D\Kstarz$~\cite{LHCb-PAPER-2016-006,LHCb-PAPER-2016-007} to the whole $\Bz \to D\Kp\pim$ Dalitz plot.

The LHCb Run~I data sample consists of $1 \invfb$ collected at $pp$ centre-of-mass energy $\sqrt{s} = 7 \tev$ and $2 \invfb$ at $\sqrt{s} = 8 \tev$.
The total Run~II data sample is expected to include an additional $5 \invfb$ collected at $\sqrt{s} = 13 \tev$, with the remainder of the $50 \invfb$ sample expected to be collected at $\sqrt{s} = 14 \tev$ and with trigger efficiency improved by around a factor of 2~\cite{LHCb-PAPER-2012-031}.
The expected yields after Run~II and after collecting $50 \invfb$ are estimated accounting for the known variation of the production cross-section of $B$ mesons within the LHCb acceptance up to $\sqrt{s} = 13 \tev$~\cite{LHCb-PAPER-2015-037}, and assuming linear scaling to $14 \tev$.

\begin{table}[!htb]
  \centering
  \renewcommand{\arraystretch}{1.1}
  \caption{
    Samples sizes of $\Bz \to D\Kp\pim$ decays in different $D$ final states observed or expected, according to the baseline amplitude model, in the LHCb Run~I data sample, with extrapolations to the samples that will be available after Run~II and after collecting $50 \invfb$.
  }
  \label{tab:samples}
  \begin{tabular}{crrr}
    \hline
    $D$ decay mode & Run~I & Run~I+II & $50 \invfb$ \\ 
    \hline
    $\Kp\pim$ & 2\,240 & 9\,200 & 140\,000 \\
    $\Km\pip$ & 220 & 900 & 14\,000 \\
    $\Kp\Km$ & 270 & 1\,100 & 17\,000 \\
    $\pip\pim$ & 130 & 540 & 8\,500 \\
    $\KS\pip\pim$ & 420 & 1\,700 & 27\,000 \\
    \hline
  \end{tabular}
\end{table}

It is of prime interest to investigate the optimal binning of the $B$ decay Dalitz plot, though certain other variations of the conditions are also considered as discussed below.
It has previously been shown~\cite{Bondar:2008hh,Gershon:2009qr} that optimising a ``binning quality factor'' $Q^2$ leads to good sensitivity to $\gamma$.
In the limit of zero background, $Q^2$ is related to the sensitivity to the interference term between suppressed and favoured amplitudes in Eq.~(\ref{eq:Adbl-Dlz}) and can be expressed as 
\begin{equation}
  \label{eq:binning-quality-simple}
  Q^2=\frac{\sum_{\alpha}\kappa_{\alpha}(\varkappa^2_{\alpha}+\sigma^2_{\alpha})}{\sum_{\alpha}\kappa_{\alpha}}\,. 
\end{equation}
The binning that maximises this expression for $Q^2$ is obtained by a stochastic optimisation procedure described in detail in Ref.~\cite{Libby:2010nu}. 

\begin{figure}[htb!]
  \centering
  \includegraphics[width=0.47\textwidth]{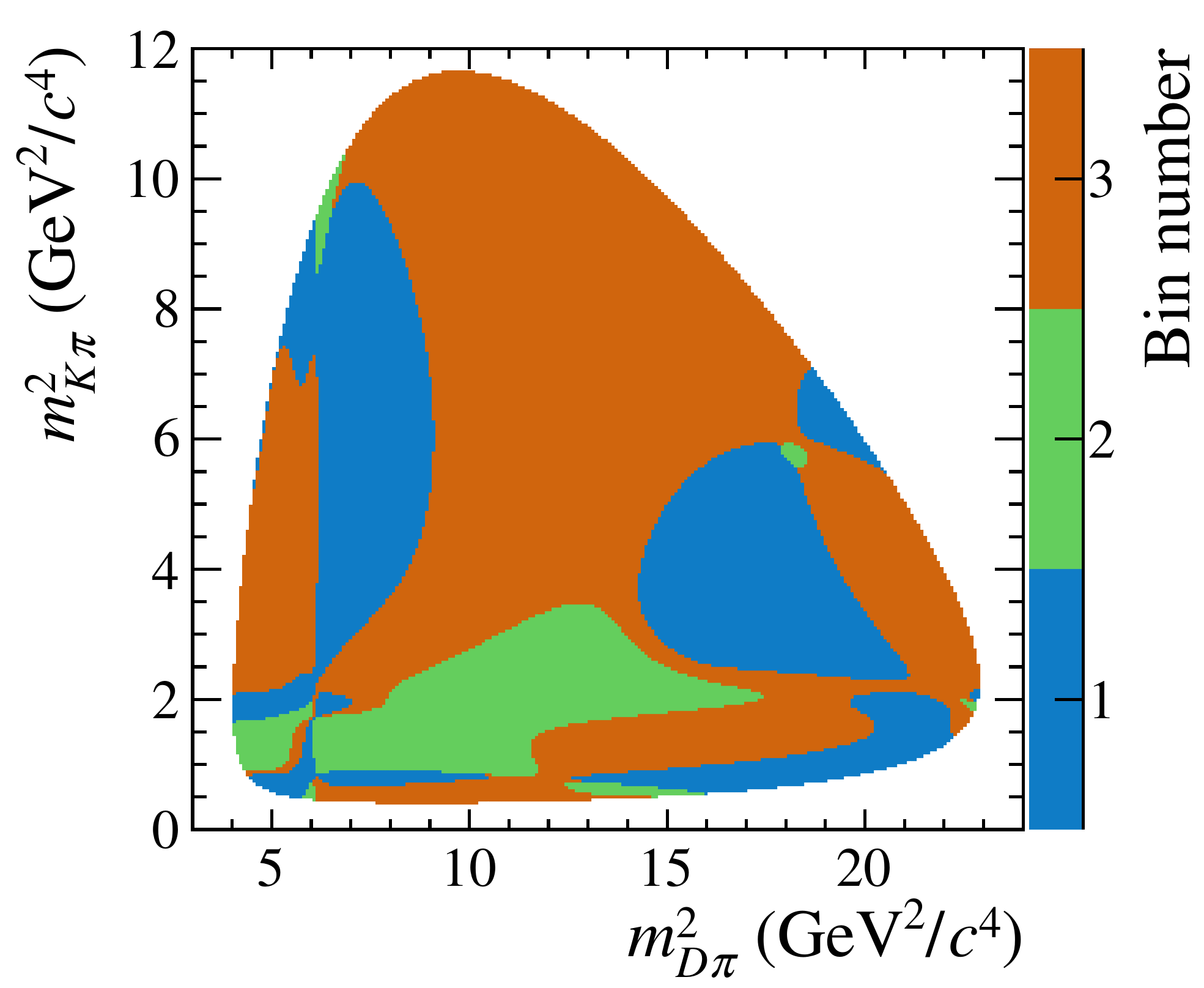}
  \hfill
  \includegraphics[width=0.47\textwidth]{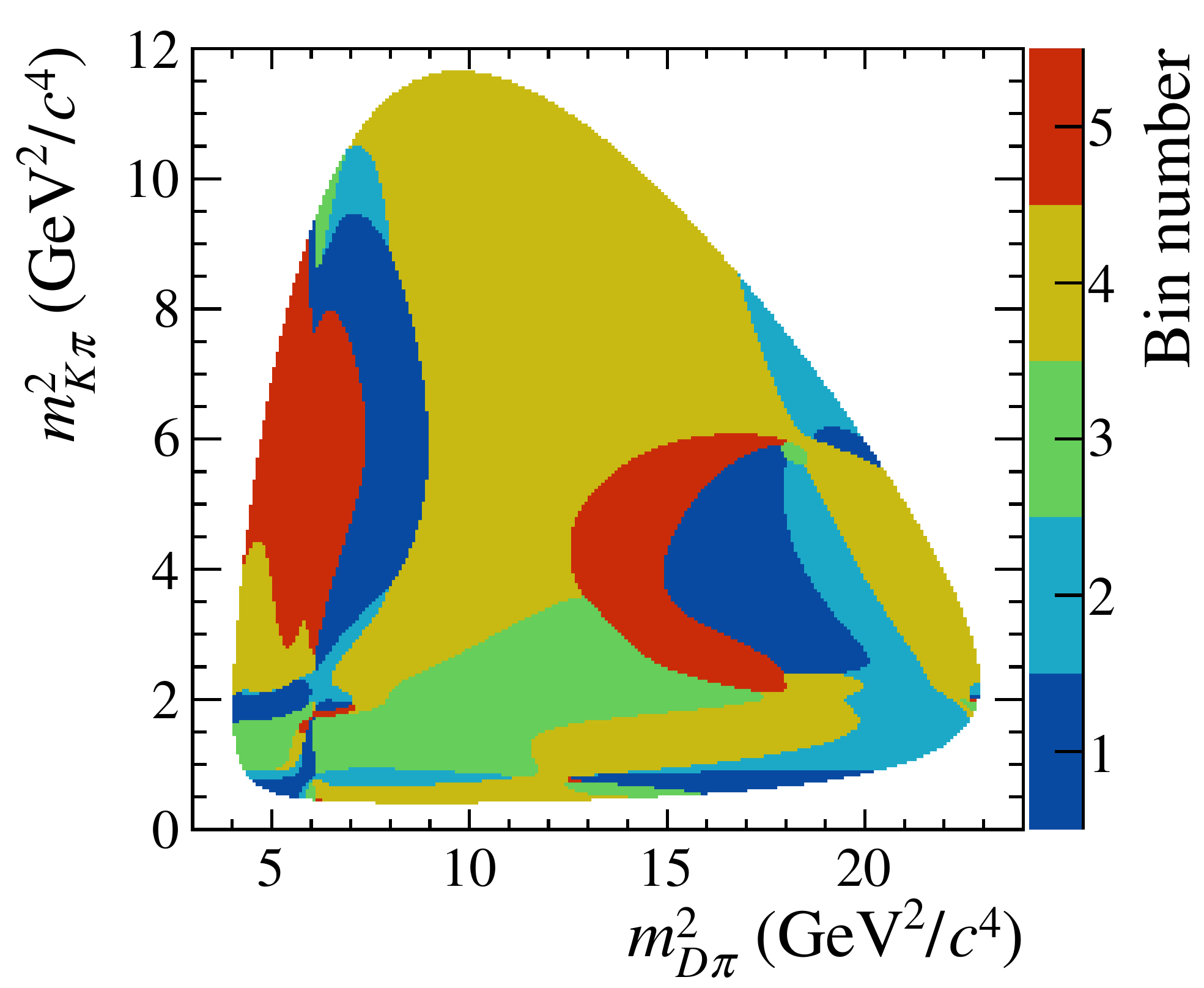}

  \includegraphics[width=0.47\textwidth]{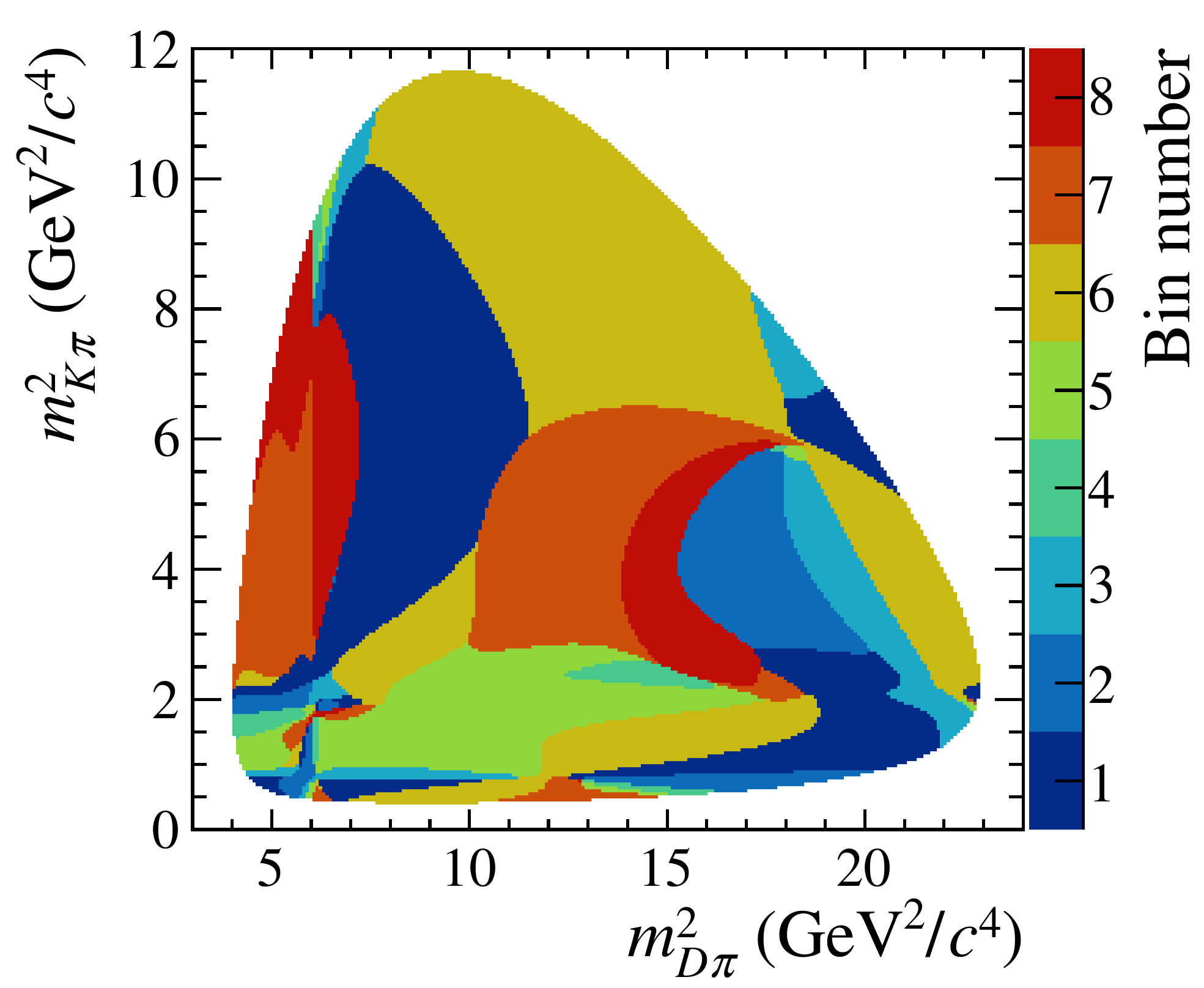}
  \hfill
  \includegraphics[width=0.47\textwidth]{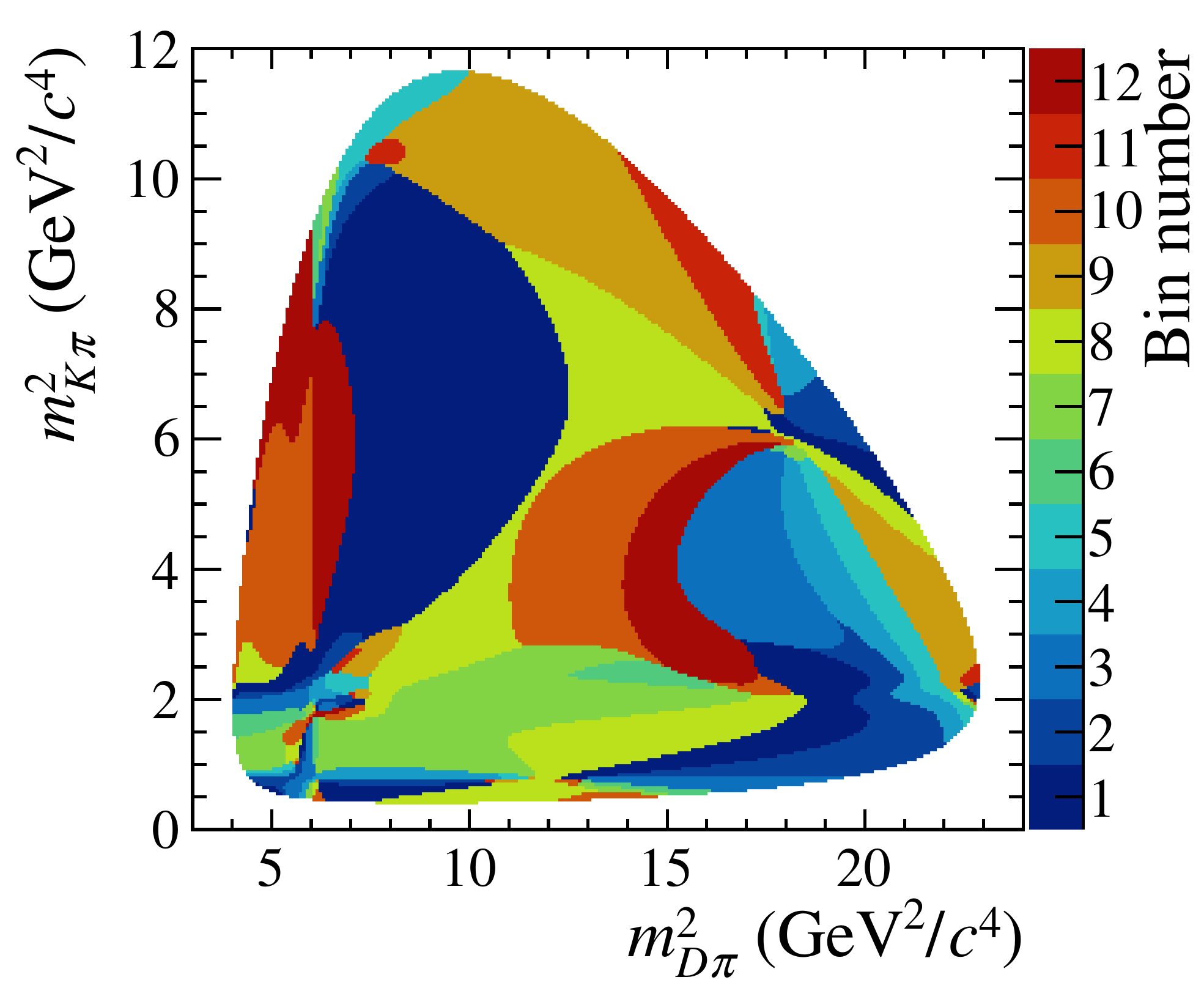}

  \includegraphics[width=0.47\textwidth]{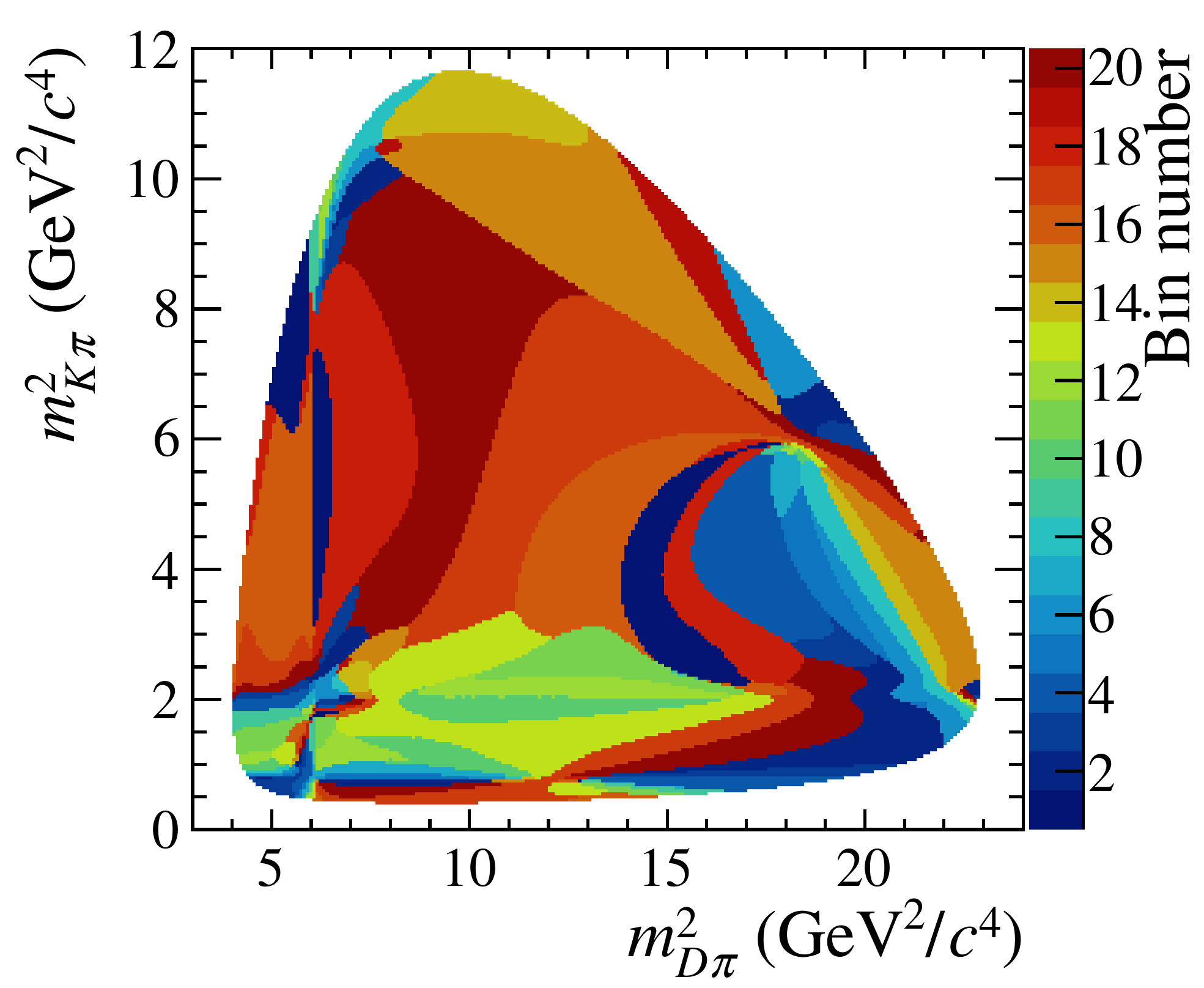}

  \caption{
    Different $B$ Dalitz plot binning schemes obtained by maximising $Q^2$ with $\mathcal{N} = 3, 5, 8, 12$ and $20$, as indicated by the $z$-axis.
  }
  \label{fig:DP-binning}
\end{figure}

Schemes with different numbers of bins in the $B$ decay Dalitz plot are considered.
Examples of the binning obtained by maximising $Q^2$ with the baseline amplitude model for $\mathcal{N} = 3, 5, 8, 12$, 
and $20$ are shown in Fig.~\ref{fig:DP-binning}.

The $D \to \KS\pip\pim$ Dalitz plot is binned with $\mathcal{N} = 8$, which has become the de~facto standard in the literature.
The values of $c_i$ and $s_i$ are calculated from the $D$ Dalitz plot amplitude model~\cite{Poluektov:2010wz}, and are consistent with those measured by the CLEO-c collaboration~\cite{Libby:2010nu}.
The effect on the sensitivity to $\gamma$ from uncertainties on $c_i$ and $s_i$ is evaluated by considering cases where $c_i$ and $s_i$ are fixed (which is the baseline), where uncertainties from Ref.~\cite{Libby:2010nu} are included as Gaussian constraints, and where $c_i$ and $s_i$ are freely floated in the fit.
The values of $K_i$ are assumed to be known with negligible uncertainty.

Only the $D$ decays to $\KS\pip\pim$ and two-body final states are included in the study, since these are expected to be the most sensitive to $\gamma$, although other channels can be added as discussed in Sec.~\ref{sec:formalism}.
The suppressed $D \to \Km\pip$ channel is expected to be the most challenging experimentally, due to the large background from $\Bs \to \Dstar \Km\pip$ decays.
Therefore, the impact of including this channel or not in the analysis is investigated.

\begin{figure}[htb!]
  \centering
  \includegraphics[width=0.47\textwidth]{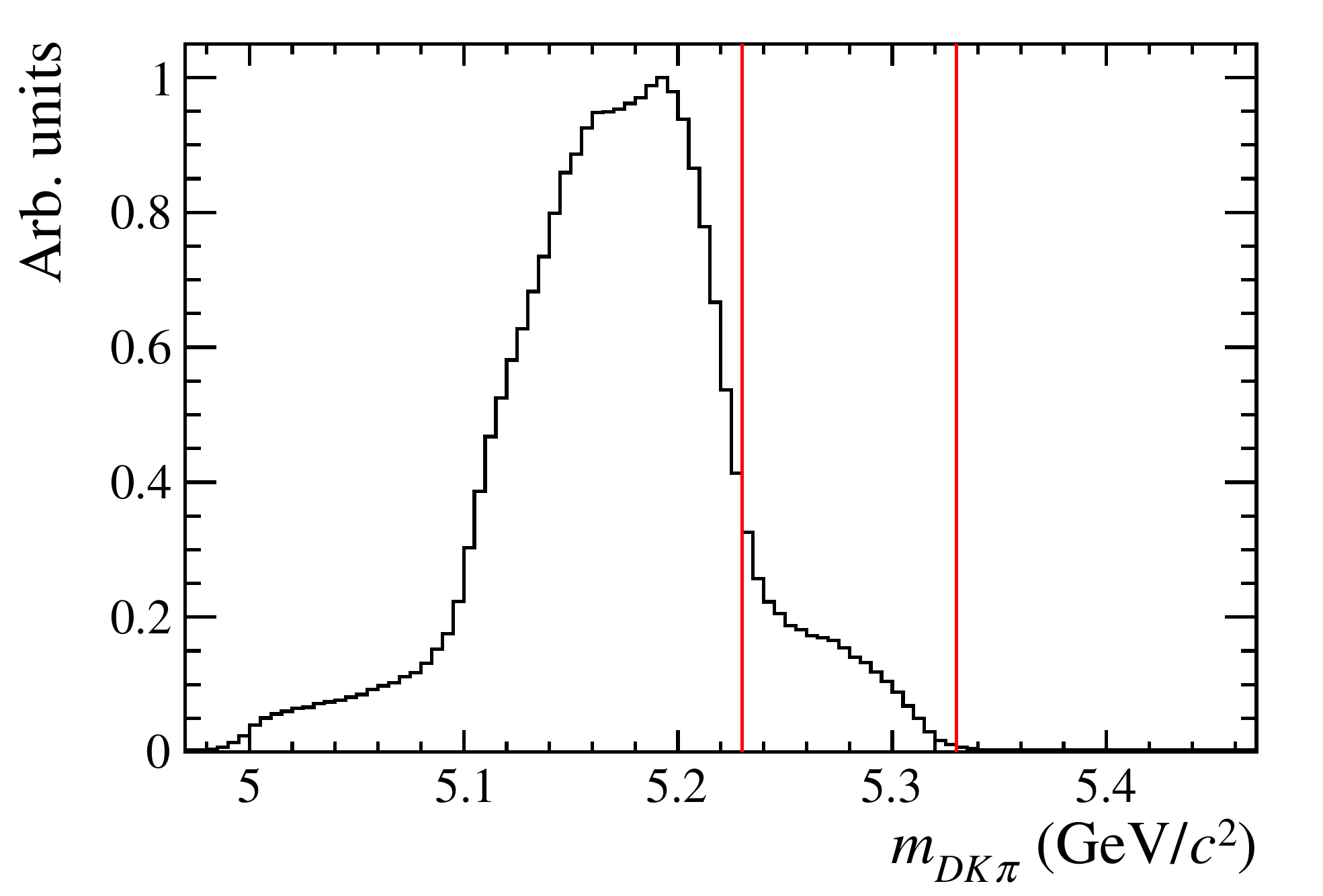}
  \put(-35, 120){(a)}
  \hspace{0.05\textwidth}
  \includegraphics[width=0.47\textwidth]{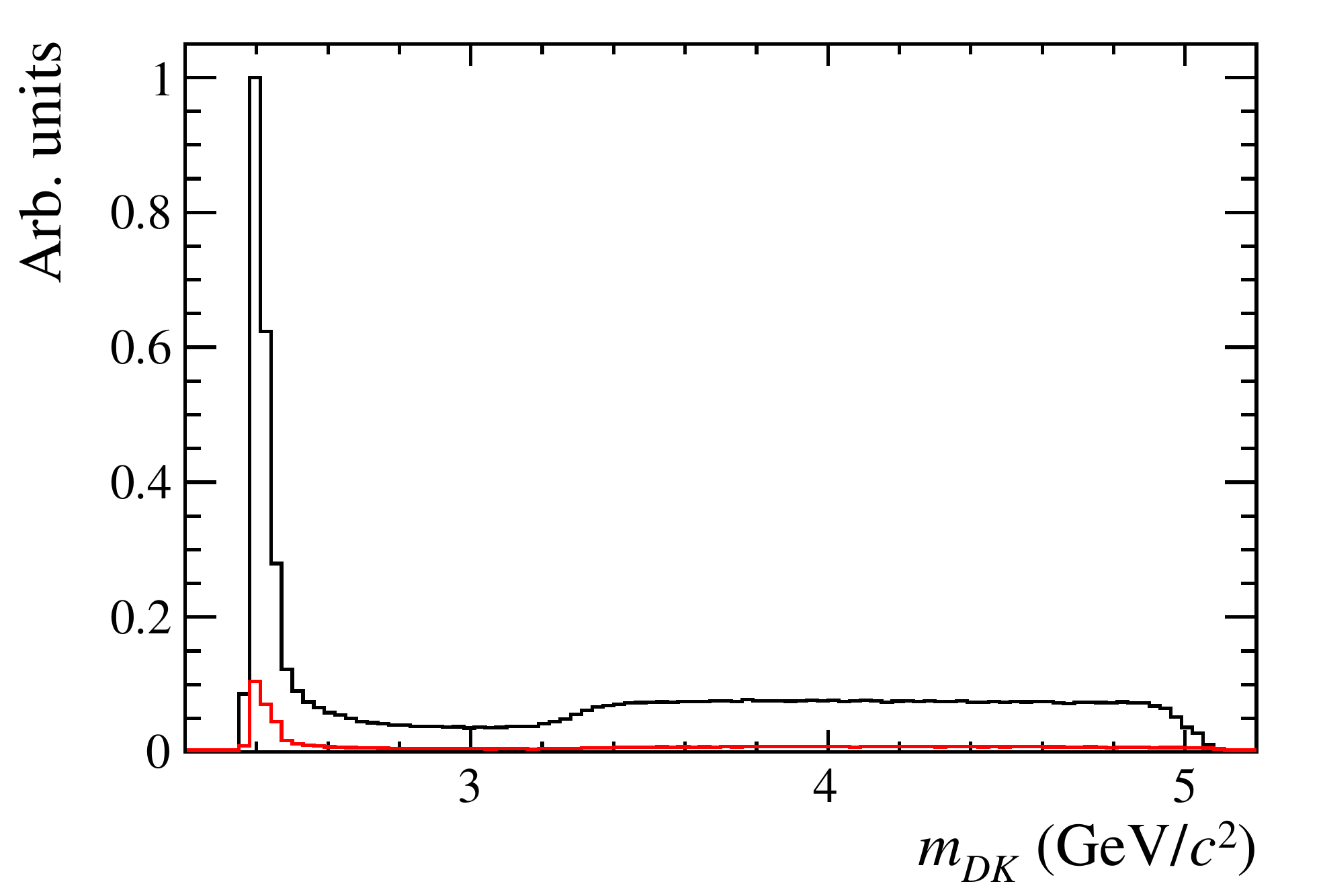}
  \put(-35, 120){(b)}

  \includegraphics[width=0.47\textwidth]{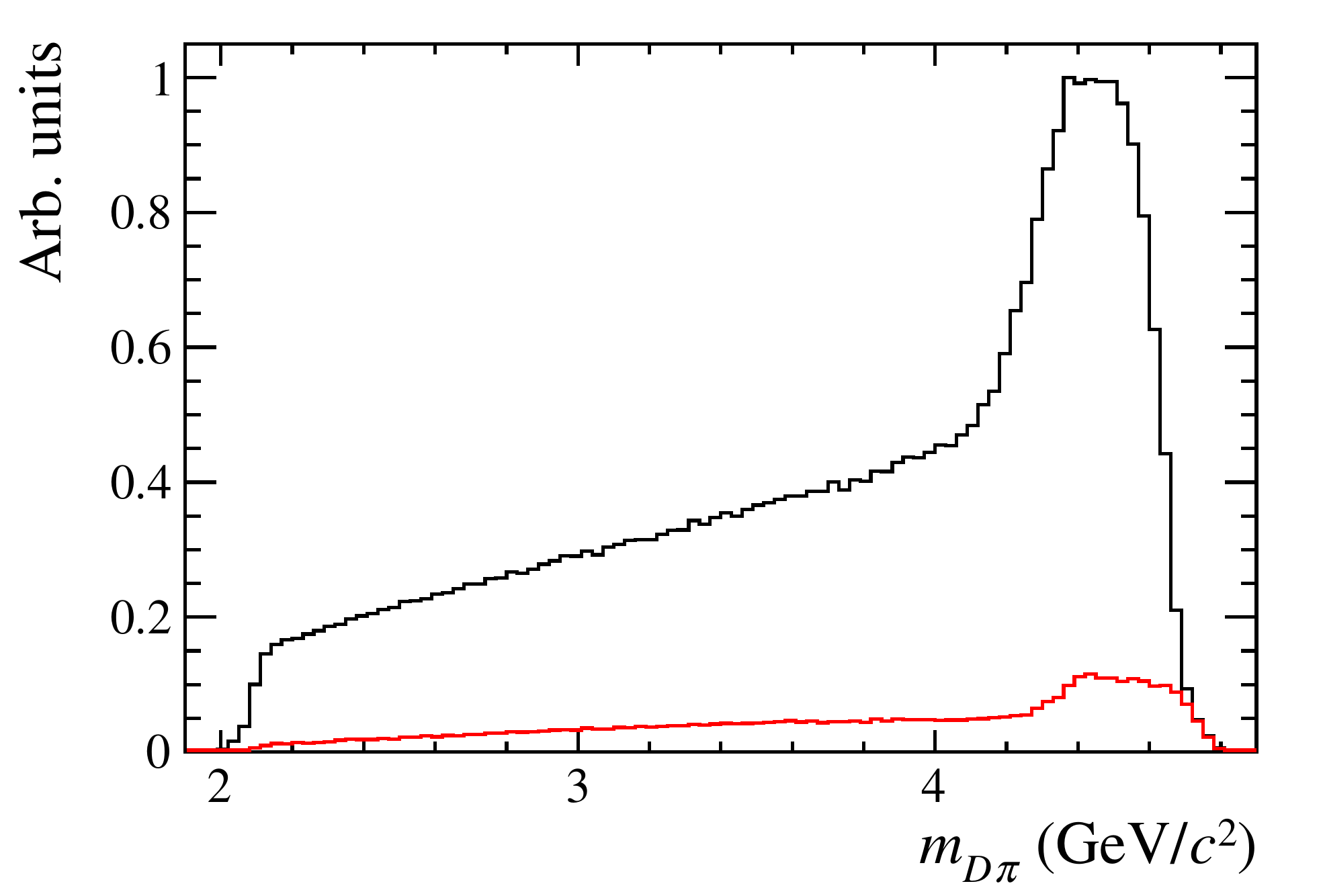}
  \put(-170, 120){(c)}
  \hspace{0.05\textwidth}
  \includegraphics[width=0.47\textwidth]{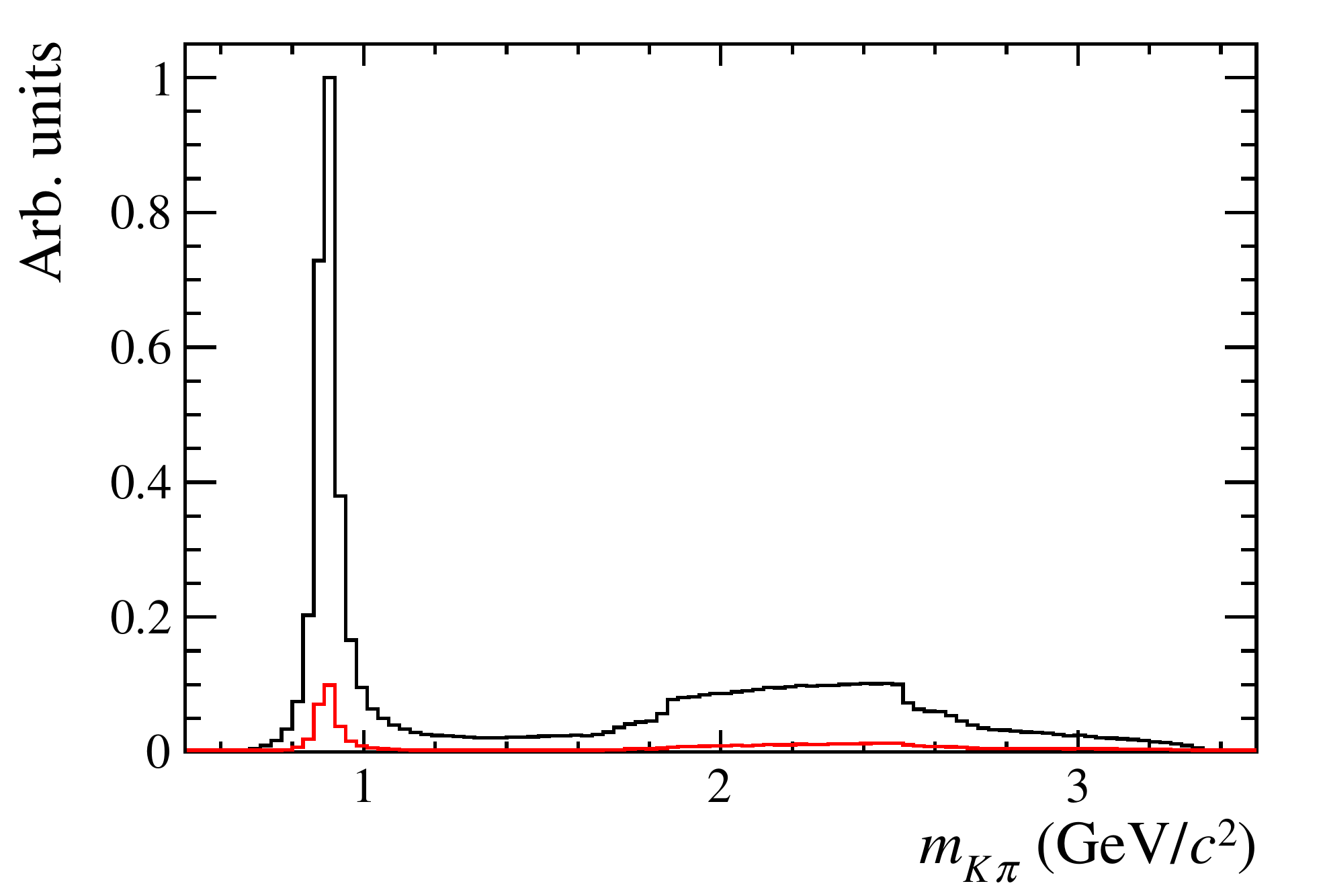}
  \put(-35, 120){(d)}
  \caption{
    Distribution of simulated $\Bs\to\Dstar\Km\pip$ decays in (top left -- bottom right) $m(D\Km\pip)$, $m(D\Km)$, $m(D\pip)$ and $m(\Km\pip)$. 
    Red histograms show the distribution of decays with $D\Km\pip$ mass within $\pm50\mevcc$ of the $\Bz$ mass, indicated by vertical red lines on the $m(D\Km\pip)$ distribution.
  }
  \label{fig:dstkpi1}
\end{figure}

\begin{figure}[htb]
  \centering
  \includegraphics[width=0.47\textwidth]{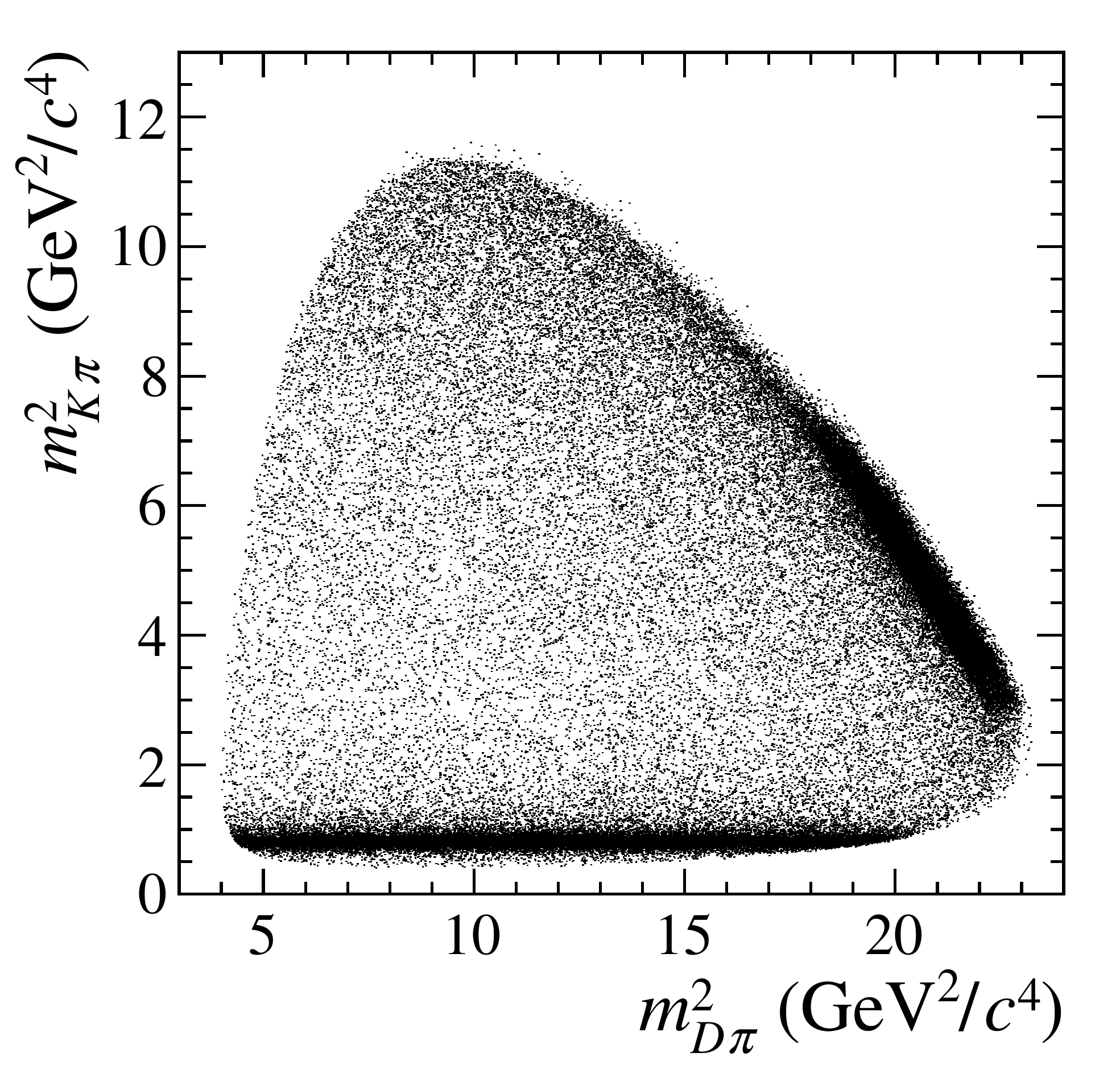}
  \caption{
    Distribution of simulated $\Bs\to\Dstar\Km\pip$ decays in the $D\Km\pip$ Dalitz plot for decays with $D\Km\pip$ invariant mass within $\pm50\mevcc$ of the $\Bz$ mass.
  }
  \label{fig:dstkpi2}
\end{figure}

The purpose of the study is to investigate the potential sensitivity, and therefore experimental effects such as backgrounds and efficiency variations are not studied.  
An exception is made for the $\Bs \to \Dstar \Km\pip$ background, which is expected to be particularly important for analyses at LHCb.
Since the amplitude structure of this decay has not yet been studied, it is modelled with a cocktail of different resonant contributions: 
$\Dstar\Kstarzb$ (45\%), $D_{s1}(2536)^-\pip$ (12\%), $D_{s2}^*(2573)^-\pip$ (12\%), $D_{s1}^*(2700)^-\pip$ (12\%) and nonresonant $\Dstar \Km\pip$ decays (7\%).
A contribution from $\Bs \to D_s(2650)^-\pip$ (12\%) is also included, where the (unobserved) $D_s(2650)^-$ state is the radially excited pseudoscalar of the charm-strange meson spectrum.
Each component of the $\Bs \to \Dstar \Km\pip$ cocktail is generated using {\tt RapidSim}~\cite{Cowan:2016tnm} and {\tt EvtGen}~\cite{Lange:2001uf}.
The distribution of the $DK\pi$ invariant mass for generated decays is shown in Fig.~\ref{fig:dstkpi1} along with distributions of the two-body invariant masses. 
The soft neutral particle from $\Dstar$ decay is not included in the reconstruction of the candidate leading to a broad $DK\pi$ invariant mass distribution peaking near $m_{\Bs} - \left( m_{\Dstarz} - m_{\Dz} \right) \approx 5.2 \gevcc$, with a significant component within the \Bz\ signal region.
Figure~\ref{fig:dstkpi2} shows the distribution of this simulated background in the $D\Km\pip$ Dalitz plot, for decays with $D\Km\pip$ invariant mass within $\pm50\mevcc$ of the \Bz mass.

In addition to LHCb, large yields of the $\Bz \to D\Kp\pim$ decay are also expected at the Belle~II experiment, which is planned to collect $50 \invab$ of $\epem$ collision data.
There is not sufficient information publicly available to make reliable estimates of the yields that can be obtained at Belle~II, and therefore this is not attempted.
Compared to LHCb, one might expect the relative yield of $D \to \KS\pip\pim$ compared to the two-body final states to be higher at Belle~II, since a larger fraction of the \KS\ mesons decay within the region in which they are reconstructible.  
However, it is not clear from the published yields in studies of $\Bz \to D\Kstarz$ decays, with $D \to \Kmp\pim$~\cite{Negishi:2012uxa} and $\KS\pip\pim$~\cite{Negishi:2015vqa} whether this is realised in practice, as the effect of different selection requirements also impacts the relative yield.
Another notable difference between LHCb and Belle~II is that it is expected to be possible to include high-yield \CP-odd channels such as $D \to \KS\piz$ in the Belle~II analysis.  
A dedicated study would be necessary to investigate the potential sensitivity of this method with the Belle~II data sample, but as a rough estimate it is expected that the precision should be around a factor of two worse than that of LHCb with $50 \invfb$, in the scenario without $\Bs \to \Dstar \Km\pip$ background.

\subsection{\boldmath Dependence of the sensitivity to $\gamma$ on the $B$ Dalitz plot binning}

Ensembles of pseudoexperiments are generated in an unbinned way, according to the $B$ and $D$ Dalitz plot models.  
The data in each pseudoexperiment are then binned according to a given scheme, and the yields in each bin are fitted to determine the following free parameters: the values of $\kappa_{\alpha}$, $\overline{\kappa}_{\alpha}$ (which are effectively determined from the favoured $\Bz \to D\Kp\pim$, $D \to \Kp\pim$ sample), $\varkappa_{\alpha}$, $\sigma_{\alpha}$, normalisation factors for each channel and $\gamma$.
The fit maximises a likelihood obtained from Eq.~(\ref{eq:num_bin}) by allowing a Poisson distribution of the yield around the expected value in each bin.
The values of $\varkappa_{\alpha}$ and $\sigma_{\alpha}$ are constrained to lie inside the physical region $\varkappa_{\alpha}^2 + \sigma_{\alpha}^2 \leq 1$; similarly $c_i^2 + s_i^2 \leq 1$ is imposed.\footnote{
  These requirements are necessary to prevent the fit from predicting, through Eq.~(\ref{eq:num_bin}), negative yields in some bins leading to an unphysical likelihood function.
}
It may be noted from Eq.~(\ref{eq:num_bin}) that there could be potential benefit from fitting for $\cos\gamma$ and $\sin\gamma$ independently, but it appears that $\gamma$ exhibits good statistical behaviour as a free parameter of the fit, and as such it is simpler to handle it in this way.
The expected uncertainty on $\gamma$ is then obtained from the spread of values obtained from the fits to pseudoexperiments in the ensemble.

\begin{figure}[htb]
  \centering
  \includegraphics[width=0.47\textwidth]{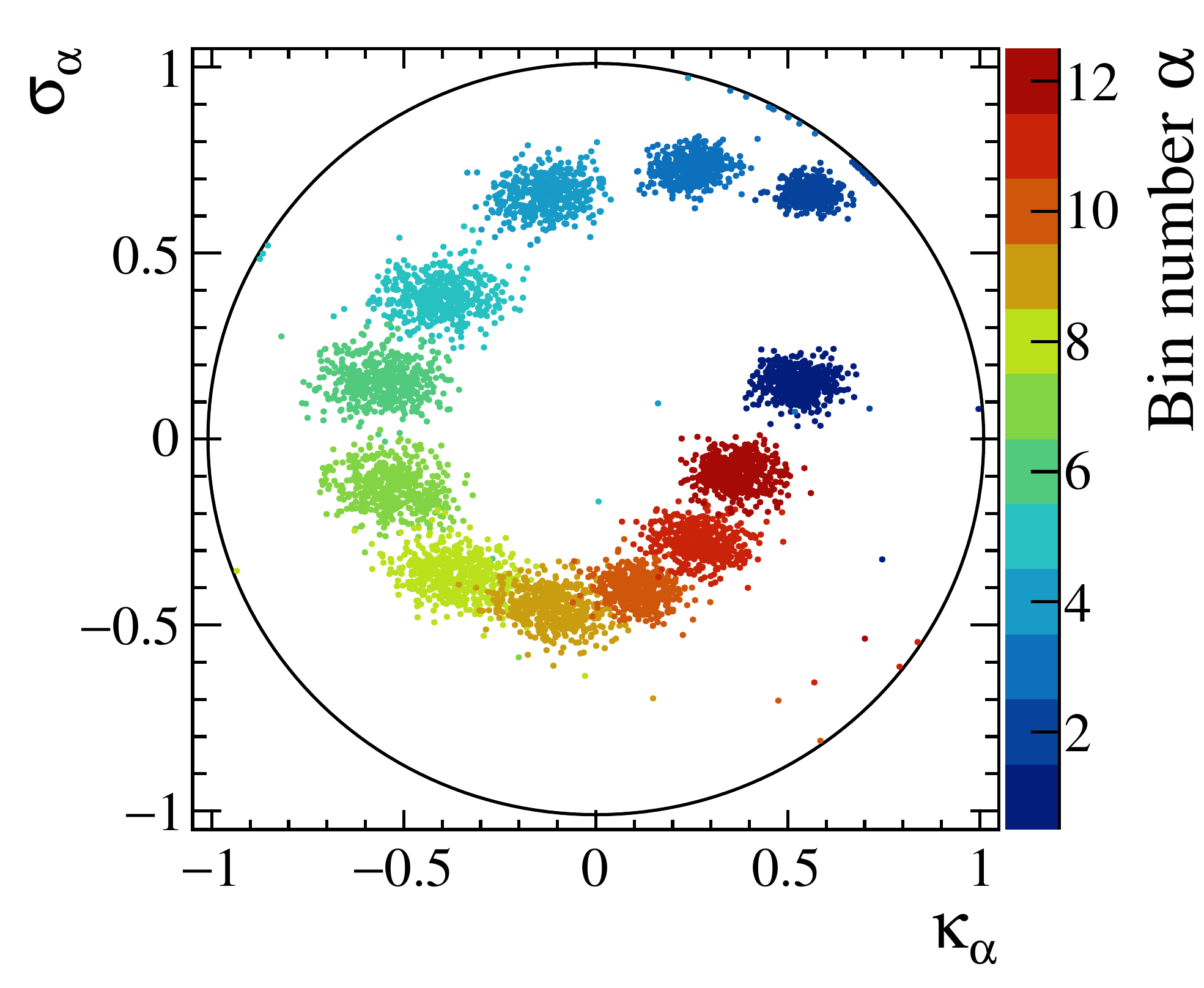}
  \put(-50, 7){\colorbox{white}{$\varkappa_{\alpha}$}}
  \put(-175,150){(a)}
  \hspace{0.03\textwidth}
  \includegraphics[width=0.47\textwidth]{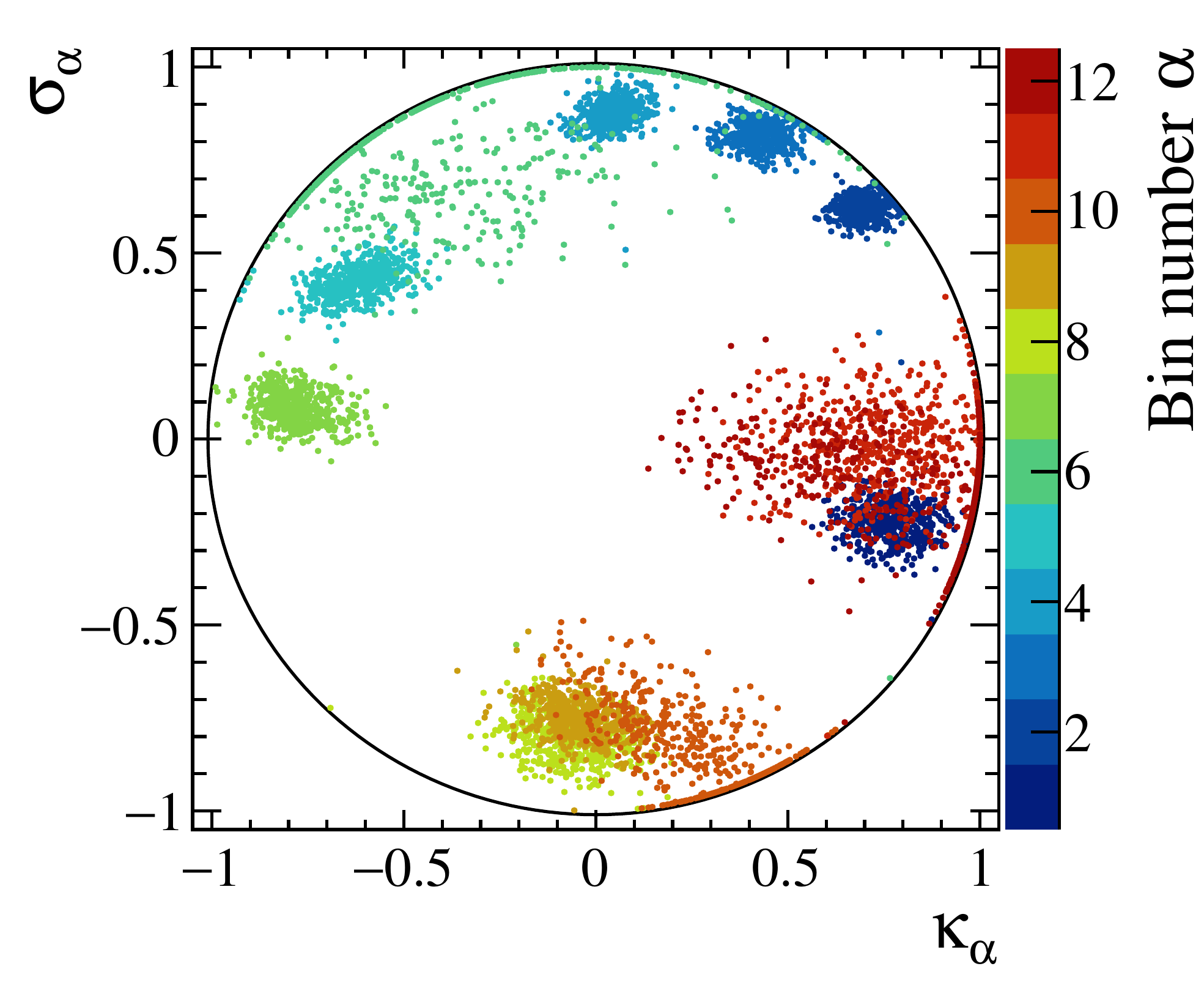}
  \put(-50, 7){\colorbox{white}{$\varkappa_{\alpha}$}}
  \put(-175,150){(b)}
  \caption{
    Fitted $\varkappa_{\alpha}$ and $\sigma_{\alpha}$ values for the $50\invfb$ scenario with 
    (a) ``equal phase-difference'' and (b) ``optimal'' binning schemes. 
  }
  \label{fig:kappa-sigma}
\end{figure}

\begin{figure}[htb]
  \centering
  \includegraphics[width=0.4\textwidth]{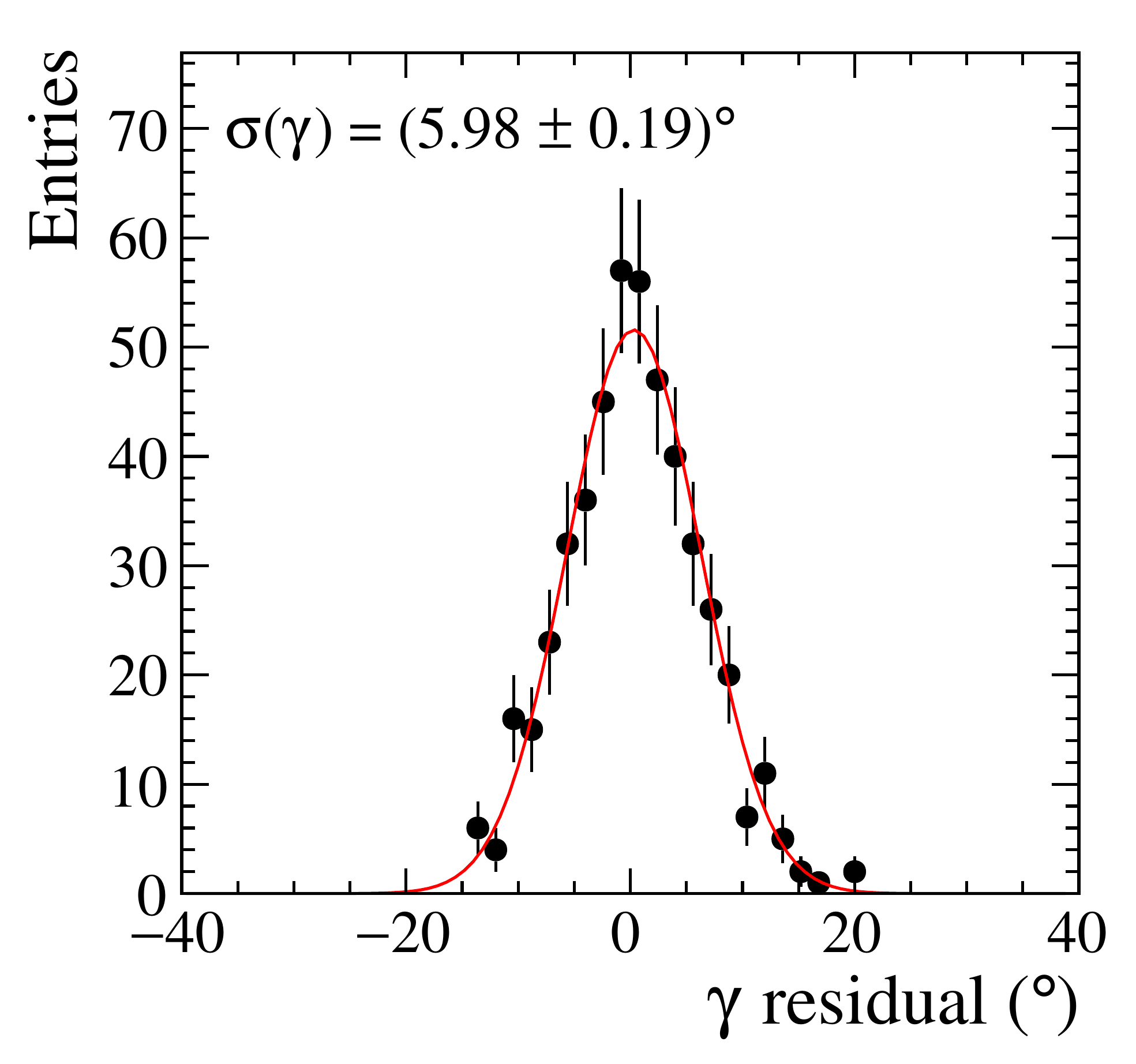}
  \put(-30,145){(a)}
  \hspace{0.03\textwidth}
  \includegraphics[width=0.4\textwidth]{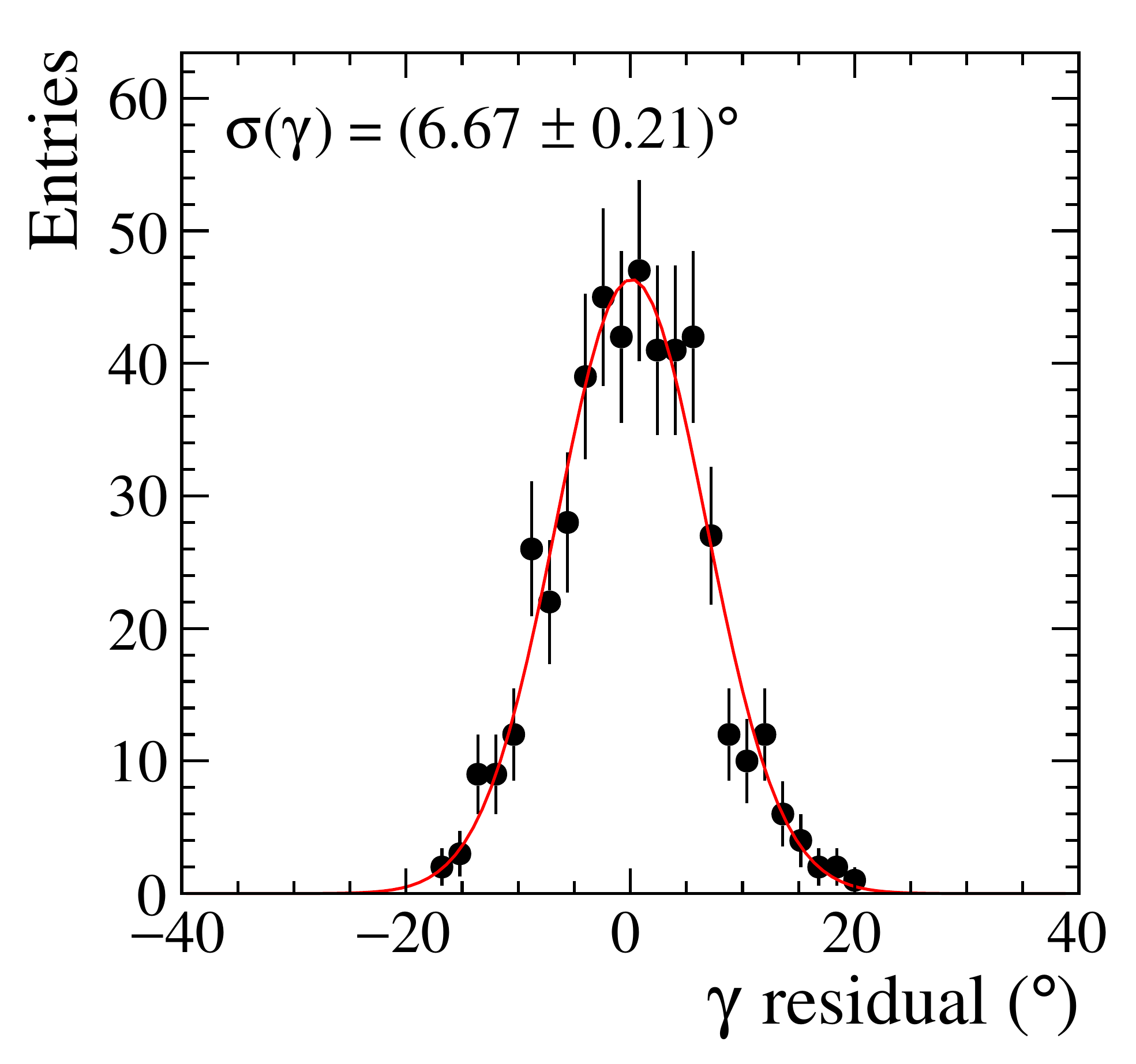}
  \put(-30,145){(b)}

  \includegraphics[width=0.4\textwidth]{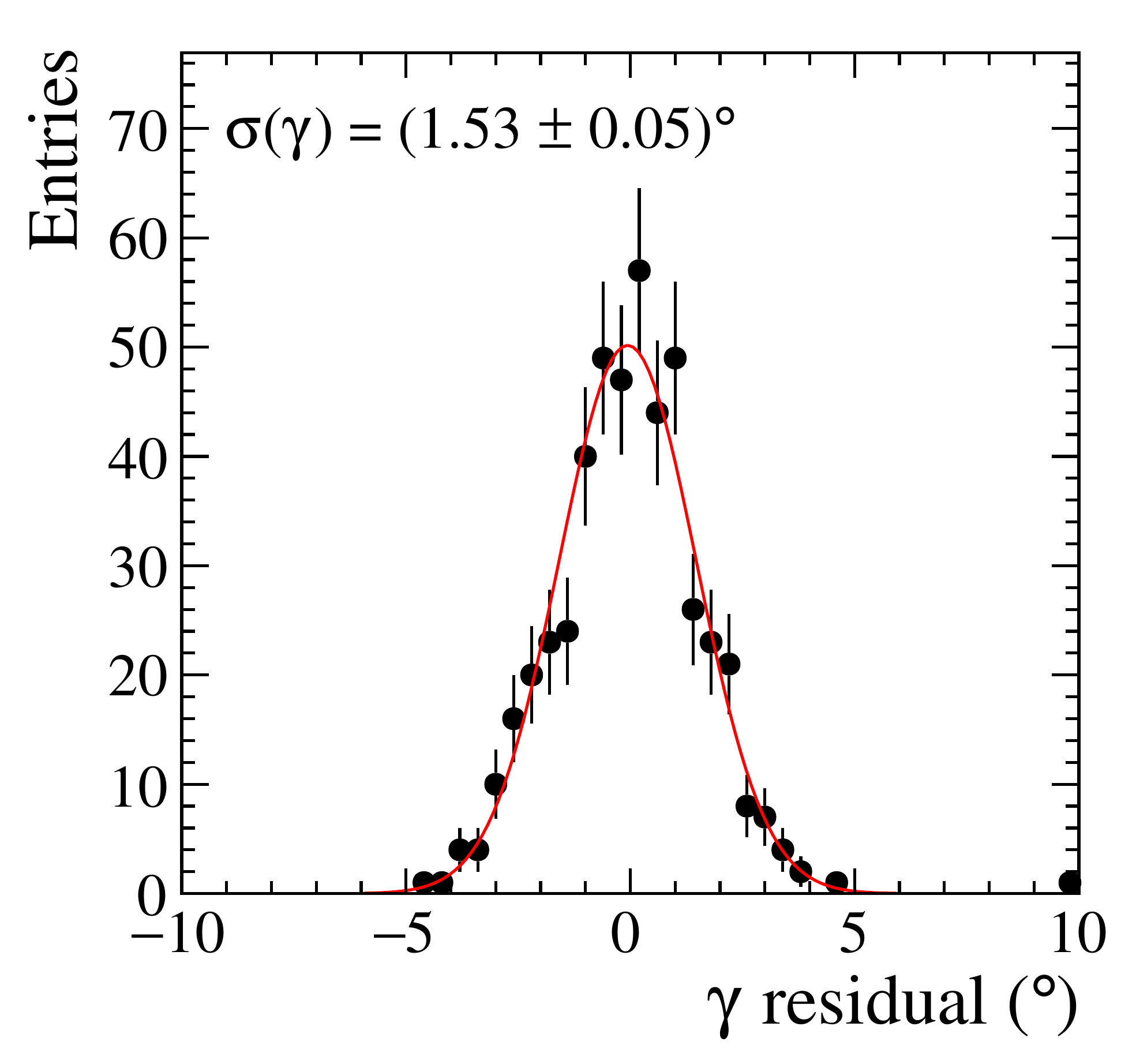}
  \put(-30,145){(c)}
  \hspace{0.03\textwidth}
  \includegraphics[width=0.4\textwidth]{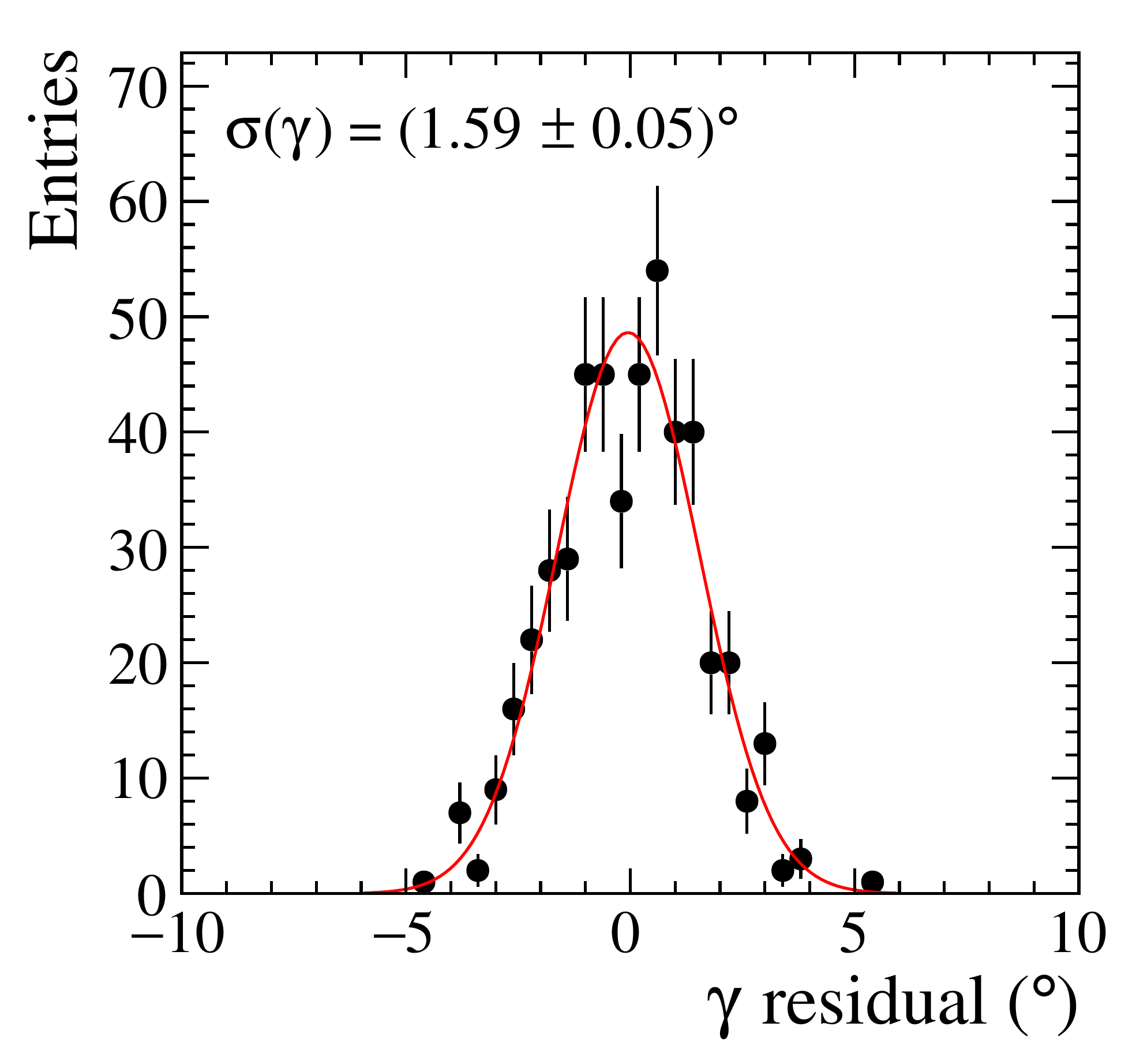}
  \put(-30,145){(d)}
  \caption{
    Residual distributions for $\gamma$ for the (a,b) Run~I+II and (c,d) $50\invfb$ scenarios (a,c) with and (b,d) without the $\Dz\to \Kp\pim$ mode. 
    The solid lines show the results of Gaussian fits to the distributions. 
  }
  \label{fig:gammaResiduals}
\end{figure}

\begin{figure}[htb]
  \centering
  \includegraphics[width=0.45\textwidth]{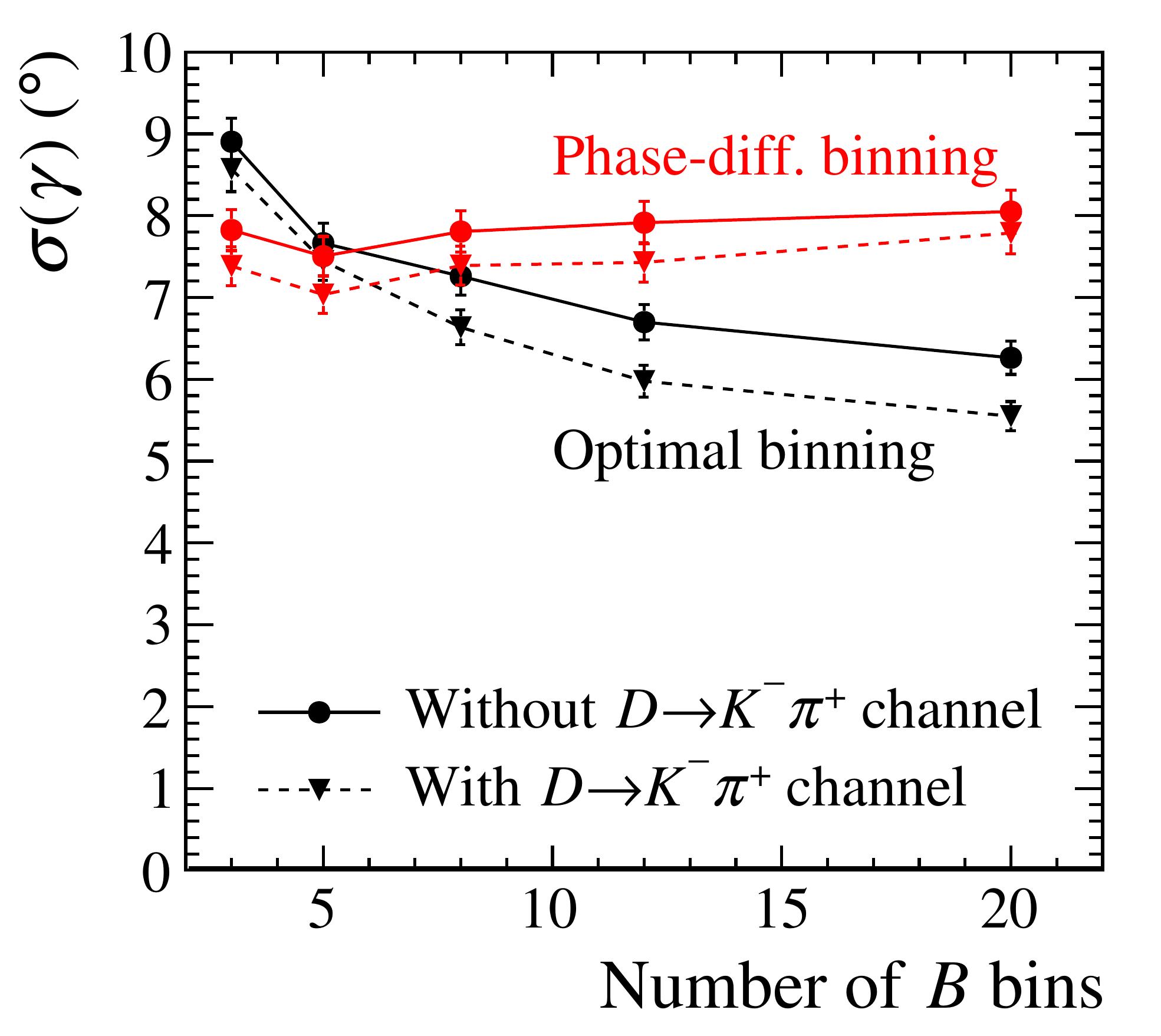}
  \put(-160,95){(a)}
  \hspace{0.05\textwidth}
  \includegraphics[width=0.45\textwidth]{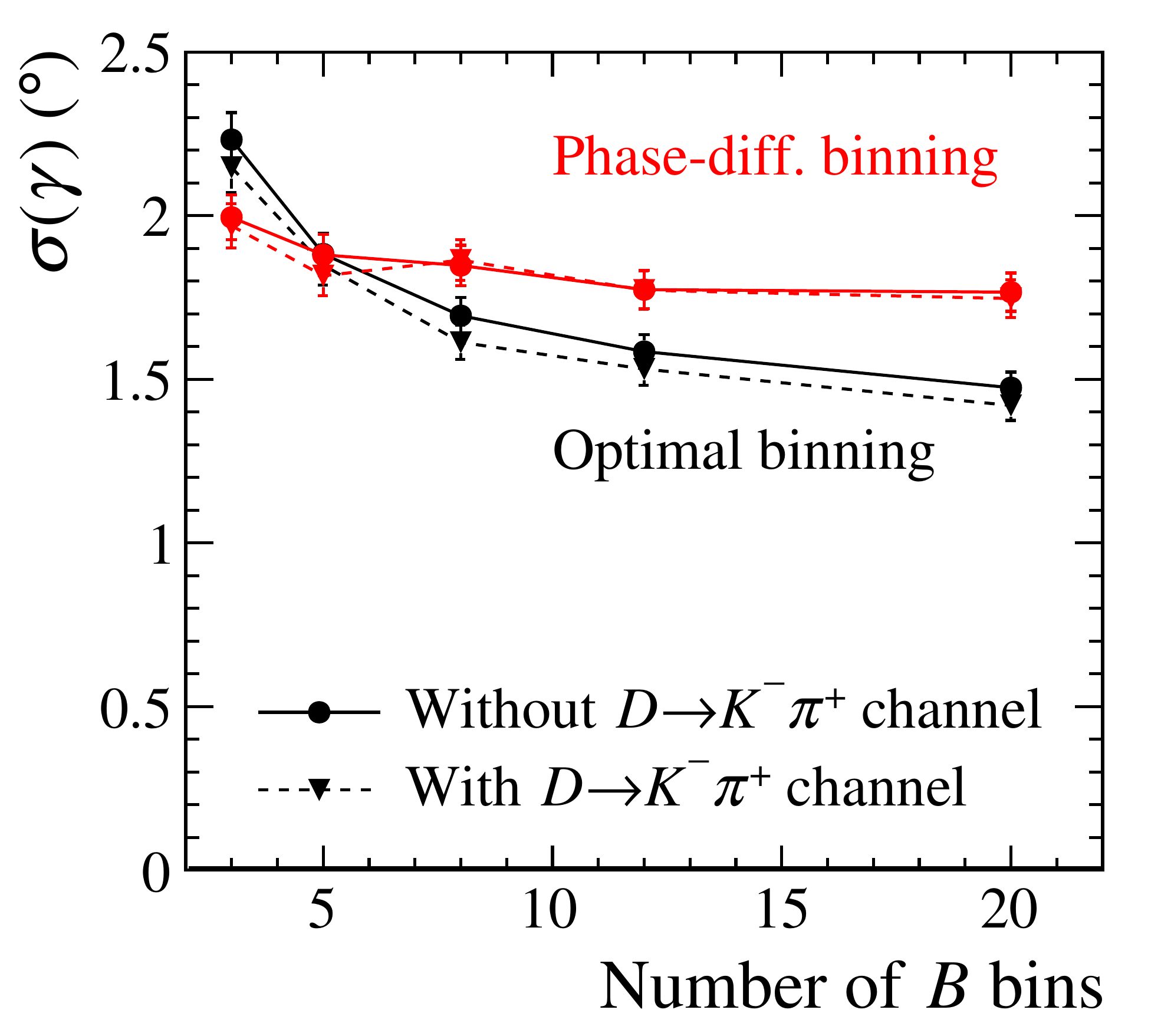}
  \put(-160,95){(b)}
  \caption{
    Sensitivity to $\gamma$ obtained with yields in each channel according to the (a)~Run~I+II and (b)~$50\invfb$ scenarios.
    Results are shown for both ``equal phase-difference'' and ``optimal'' binning schemes.
    The lines joining the points are added simply to guide the eye.
  }
  \label{fig:gammaSensitivity}
\end{figure}

The results of the fits are shown in Figs.~\ref{fig:kappa-sigma}, \ref{fig:gammaResiduals} and \ref{fig:gammaSensitivity}. 
Figure~\ref{fig:kappa-sigma} illustrates the effect of ``optimal'' binning compared to an alternative binning with uniform division of the the strong phase difference between the $\Bz \to \Dz\Kp\pim$ and $\Bz \to \Dzb\Kp\pim$ amplitudes (``equal phase-difference'' binning).
The fits are performed to samples corresponding to the $50\invfb$ scenario with the baseline amplitude model. 
The result of the fit for each pseudoexperiment is represented by a coloured point, where the colour denotes the bin number $\alpha$.
It can be seen that the ``optimal'' binning results in $\varkappa_{\alpha}$, $\sigma_{\alpha}$ values that tend to be closer to the unit circle, corresponding to higher coherence in each of the bins and thus better sensitivity according to Eq.~(\ref{eq:binning-quality-simple}). 

Figure~\ref{fig:gammaResiduals} shows residual distributions for $\gamma$ obtained from the fits, for each of the Run~I+II and $50\invfb$ scenarios both with and without the suppressed $D\to \Km\pip$ mode included in the likelihood.
Pseudoexperiments are generated with the baseline model and the fits are performed with the ``optimal'' binning with $\mathcal{M}=12$.
In all cases there is no visible bias in $\gamma$.
The absence of significant bias is also verified for all binning schemes used in subsequent fits, with $\mathcal{M}=3,5,8,12$, and $20$. 

The resolution of $\gamma$ obtained from fits to the residual distributions as a function of the number of bins $\mathcal{M}$ for both ``equal phase-difference'' and ``optimal'' binning schemes, with and without the $D\to \Kp\pim$ mode in the likelihood, are shown in Fig.~\ref{fig:gammaSensitivity}.
Overall, removing the $D\to \Kp\pim$ mode results in only 3--10\% increase in the uncertainty on $\gamma$.
The use of ``optimal'' binning results in consistently better resolution than with the ``equal phase-difference'' binning for sufficiently large number of bins ($\mathcal{M}>5$).
Therefore, ``optimal'' binning schemes are used for all subsequent studies. 

\subsection{\boldmath Dependence on the uncertainty of the $c_i$ and $s_i$ factors}

As discussed in Sec.~\ref{sec:formalism}, the coefficients $c_i$ and $s_i$ have been measured and are therefore not considered as unknown parameters.
In the baseline analysis, all $c_i$ and $s_i$ are fixed to their known true values as predicted by the $D$ decay amplitude model.
In an experimental analysis one would instead use the measured central values~\cite{Libby:2010nu}, and the values of $c_i$ and $s_i$ could be varied within their uncertainties to evaluate the associated systematic uncertainty.
However, the $\Bz\to D\Km\pip$ double Dalitz plot analysis itself also provides sensitivity to $c_i$ and $s_i$, owing to the large value of the interference term.
A natural approach is therefore to include the externally measured values of $c_i$ and $s_i$ into the likelihood with Gaussian constraints.
In this way, the uncertainty in the external determination of $c_i$ and $s_i$ enters the statistical uncertainty of the result. 
Alternatively, the $c_i$ and $s_i$ parameters can be treated as unknown and floated in the fit, removing the dependence on external measurements.

\begin{figure}[htb]
  \centering
  \includegraphics[width=0.45\textwidth]{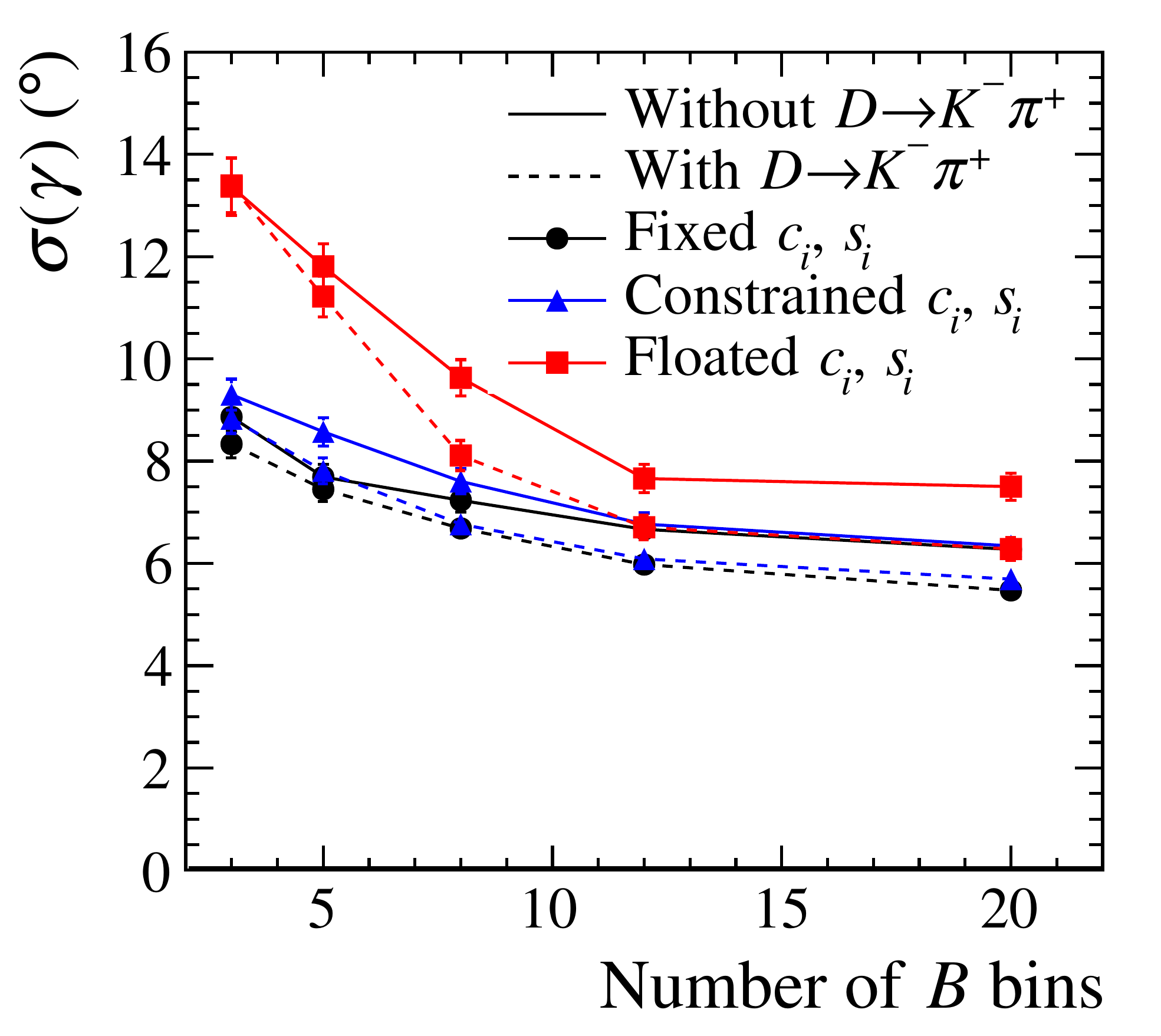}
  \put(-35,45){(a)}
  \hspace{0.05\textwidth}
  \includegraphics[width=0.45\textwidth]{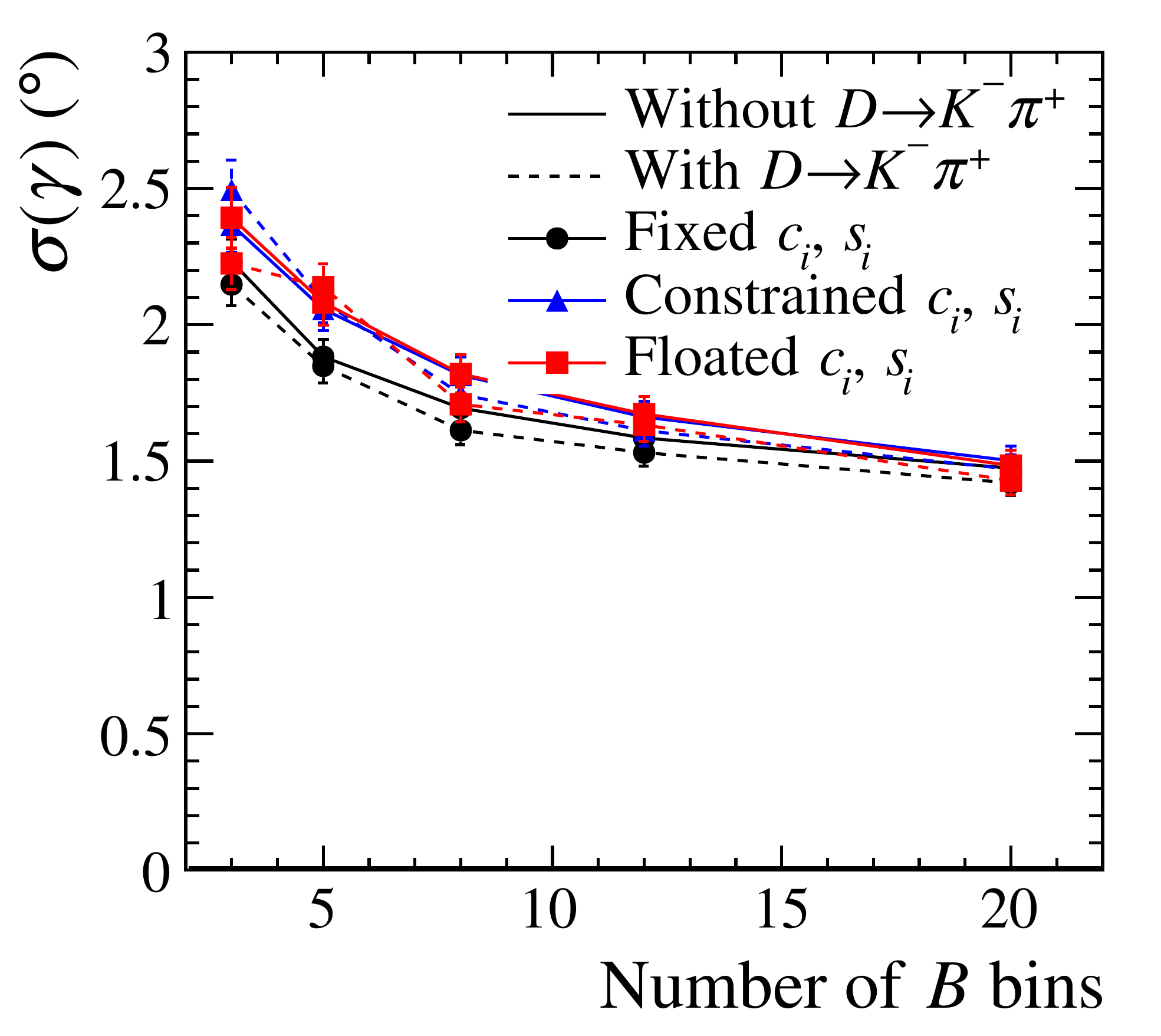}
  \put(-35,45){(b)}
  \caption{
    Impact of the uncertainty on the $c_i$ and $s_i$ parameters on the sensitivity to $\gamma$ for the (a)~Run~I+II and (b)~$50\invfb$ scenarios.
    Results are shown with $c_i, s_i$ values fixed to their true values, constrained within the precision of the current measurements~\cite{Libby:2010nu}, and left free to float in the fit. 
    The lines joining the points are added simply to guide the eye.
  }
  \label{fig:CSSensitivity}
\end{figure}

\begin{figure}[htb]
  \centering
  \includegraphics[width=0.45\textwidth]{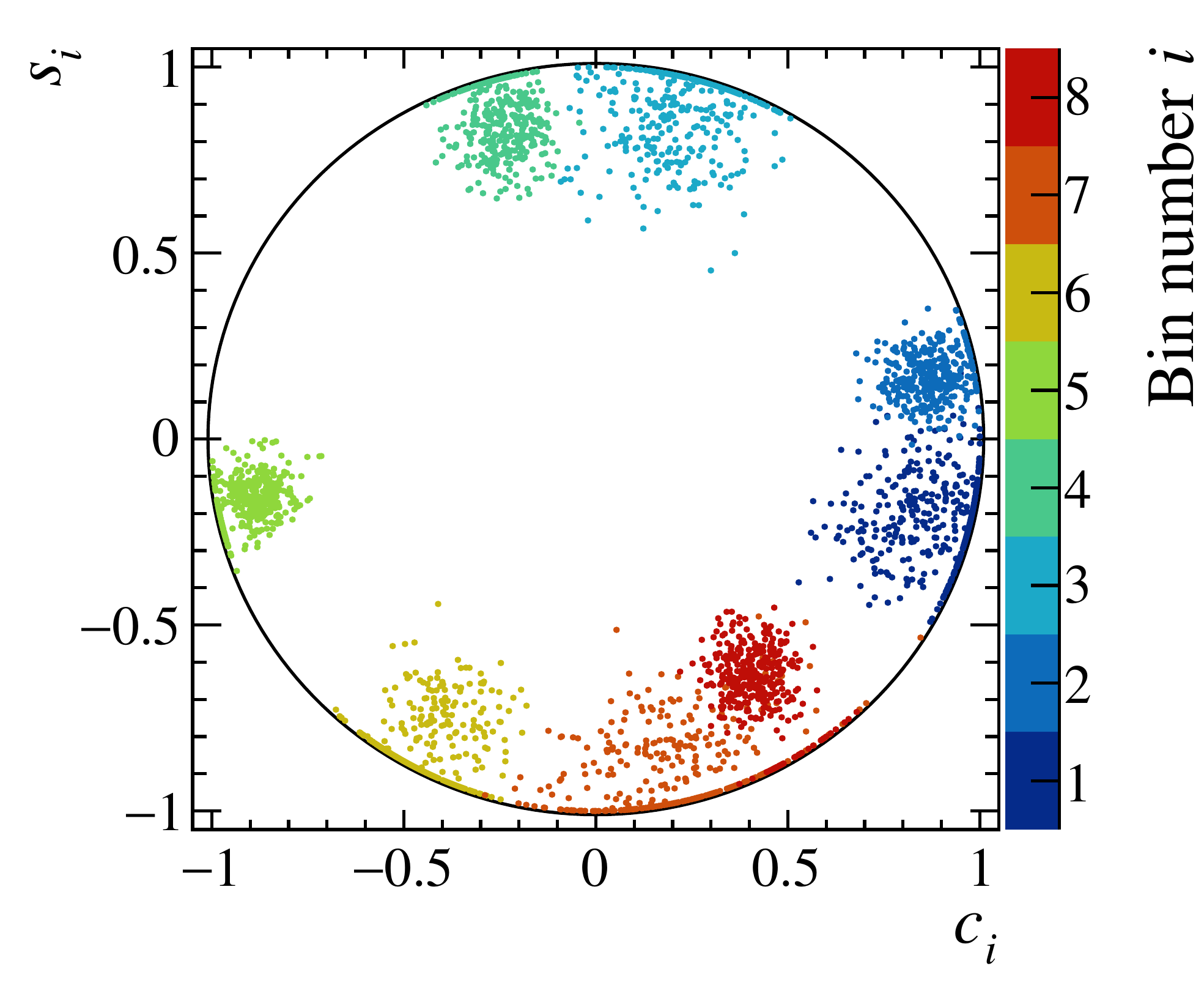}
  \caption{
    Fitted $c_i$ and $s_i$ values for the $50\invfb$ scenario with $\mathcal{M}=8$ and no external constraints.
  }
  \label{fig:floatCS}
\end{figure}

The impact of these different approaches to external constraints on $c_i$ and $s_i$ is illustrated in Fig.~\ref{fig:CSSensitivity}, which shows the resolution on $\gamma$ as a function of the number of bins $\mathcal{M}$ for the cases when $c_i$ and $s_i$ terms are fixed to their true values, when Gaussian constraints are applied corresponding to the current measurement uncertainties~\cite{Libby:2010nu}, and when $c_i$ and $s_i$ are left unconstrained. 
The difference between the extreme cases of fixing or floating the $c_i$ and $s_i$ parameters is quite significant for the Run~I+II scenario, particularly for smaller $\mathcal{M}$.
However, the precision of the current measurements of $c_i$ and $s_i$ appears to be sufficient so that the sensitivity to $\gamma$ is not degraded substantially.
Interestingly, as the data sample increases, the importance of precise external measurements of $c_i$ and $s_i$ reduces, in contrast to the situation for the model-independent analysis of $\Bp \to D\Kp$ with $D \to \KS\pip\pim$ decays~\cite{Malde:2223391}, as the double Dalitz plot analysis itself constrains these parameters.
Figure~\ref{fig:floatCS} shows as an example the fitted values of the $c_i$ and $s_i$ parameters from fits in the $50\invfb$ scenario; the uncertainties are in the range 0.07--0.17, comparable to or somewhat better than those of the current measurements~\cite{Libby:2010nu}.
Nonetheless, precise independent external measurements of $c_i$ and $s_i$, as could be obtained by the BESIII experiment, would remain important to provide a cross-check of the measurement.  

\subsection{\boldmath Dependence on the value of $r_B$}

\begin{figure}[htb]
  \centering
  \includegraphics[width=0.45\textwidth]{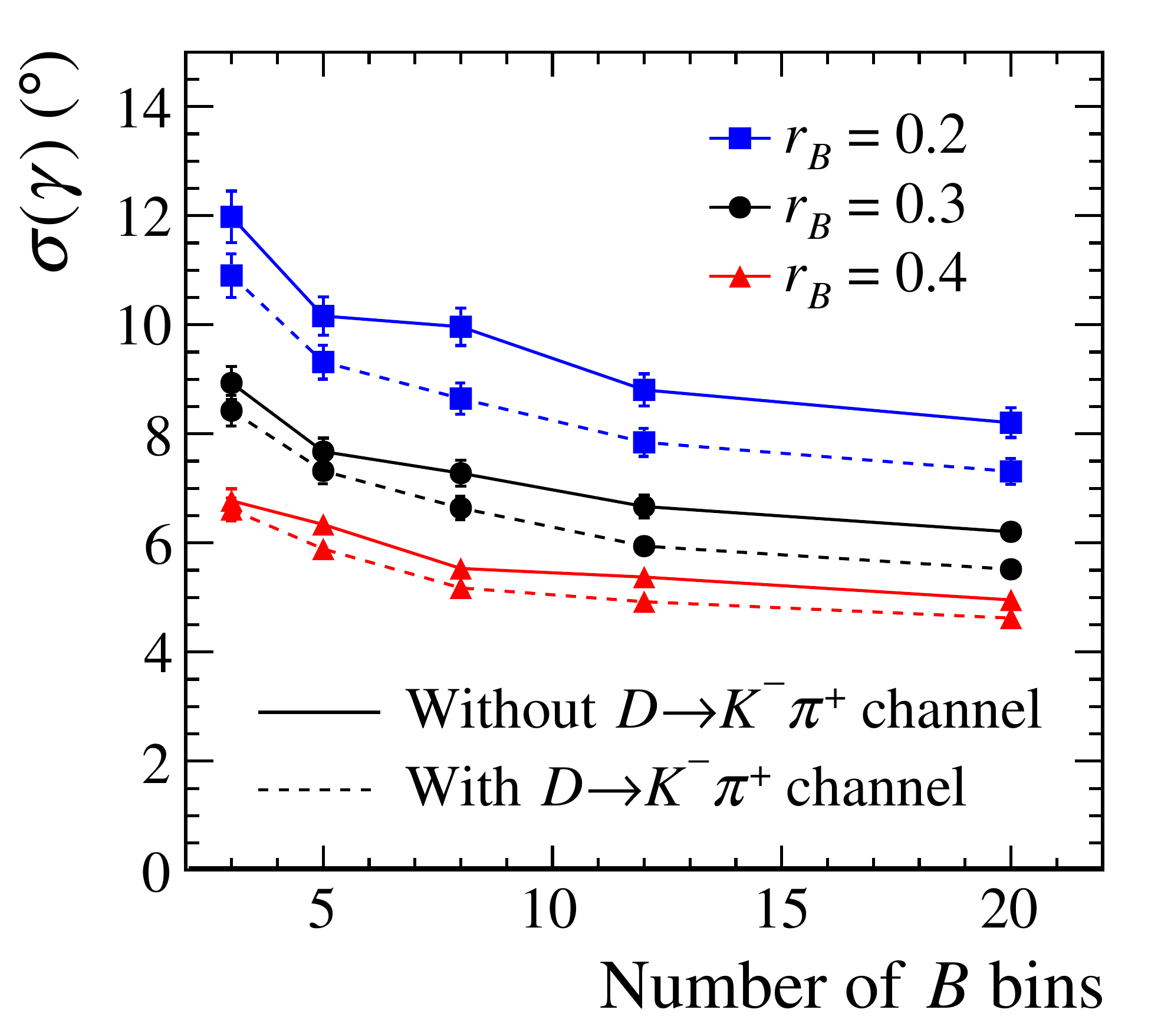}
  \put(-155, 155){(a)}
  \hspace{0.05\textwidth}
  \includegraphics[width=0.45\textwidth]{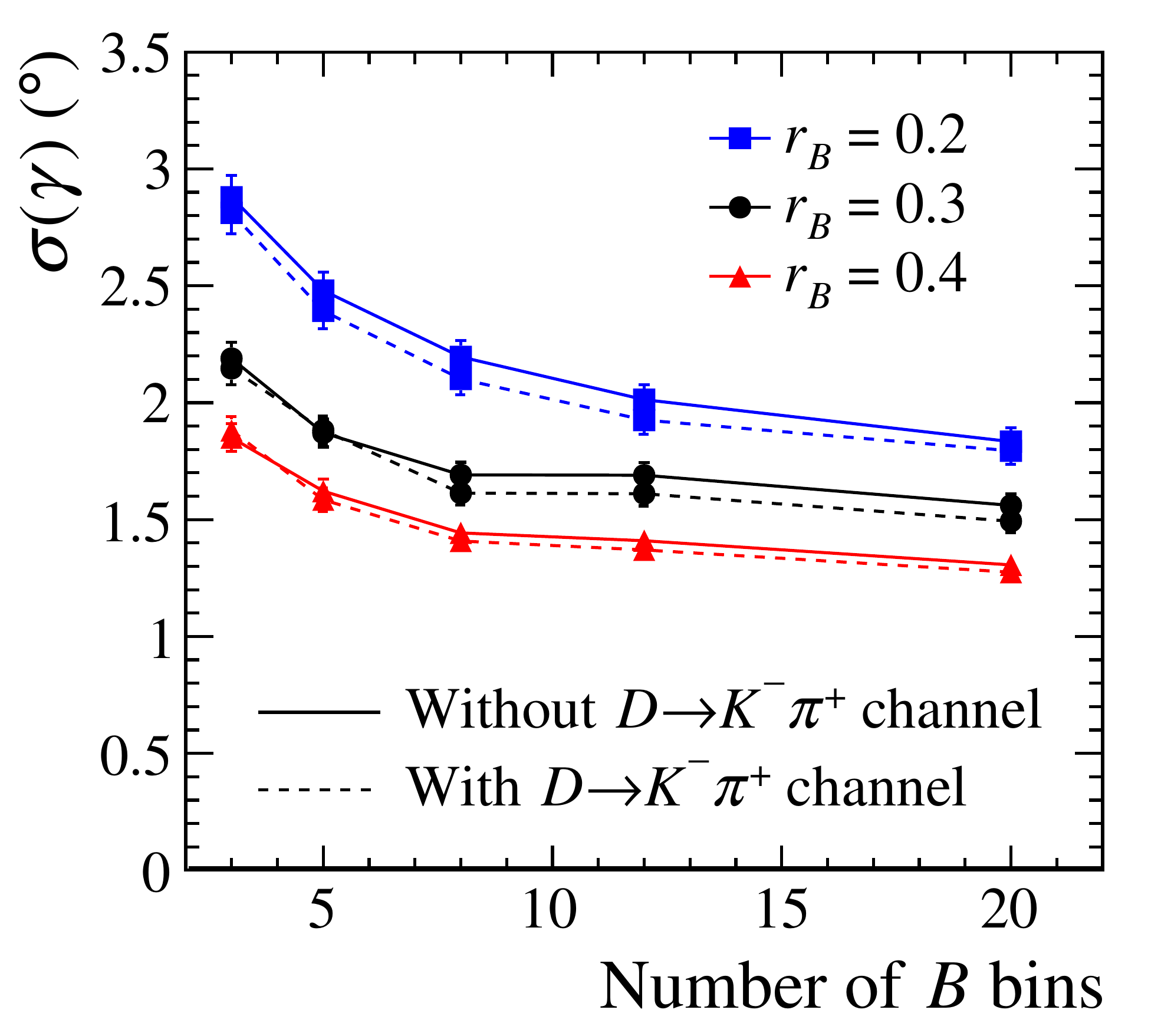}
  \put(-155, 155){(b)}
  \caption{
    Sensitivity to $\gamma$ in the (a) ``Run~I+II'' and (b) ``50\invfb'' scenarios, as a function of number of bins used for the $B$ Dalitz plot, shown for $r_B = 0.2$ (blue), $0.3$ (black) and $0.4$ (red).
    The lines joining the points are added simply to guide the eye.
  }
  \label{fig:rB-sensitivity}
\end{figure}

The sensitivity to $\gamma$ is expected to have a strong dependence on the ratio of magnitudes of the suppressed and favoured amplitudes.
For (quasi-)two-body decays, this ratio is quantified by the value $r_B$, which can differ for each kaonic state produced in a \mbox{$B \to DK$-type} process.  
In the baseline model, $r_B = 0.3$ is used for all of the $K^*(892)^0$, $K^*(1410)^0$, $K_2^*(1430)^0$ and $K\pi$ S-wave contributions.
The effect of varying $r_B$ to smaller or larger values is shown in Fig.~\ref{fig:rB-sensitivity}.
As expected, larger values of $r_B$ result in better sensitivity.
It can also be noted that the impact of the $D \to \Km\pip$ channel is more significant for smaller values of $r_B$.  

\subsection{\boldmath Effect of $\Bs \to \Dstar \Km\pip$ background}

Based on the yield of the $\Bs \to \Dstar \Km\pip$ background in Ref.~\cite{LHCb-PAPER-2015-059}, the expected level of this background relative to signal in the modes $\Bz\to D\Km\pip$ with $D\to \Kp\Km$, $\pip\pim$ and $\KS\pip\pim$ is around 20\% in the $\pm 35\mevcc$ region around the $B$ meson mass.
For the suppressed mode with $D\to \Km\pip$, however, the background-to-signal ratio is expected to be around $7.5$.
This mode is therefore considered to be background-dominated and is not considered in the background-enabled fits. 

The expected background yields $\langle N^{\rm(bck)}_{\alpha i}\rangle$ ($\langle N^{\rm(bck)}_{\alpha}\rangle$) are calculated for each $B$ bin $\alpha$ and $D$ bin $i$ for the $D\to\KS\pip\pim$ mode (for each $B$ bin $\alpha$ for two-body $D$ decays) according to the expected Dalitz plot distribution (Figs.~\ref{fig:dstkpi1} and \ref{fig:dstkpi2}) and assuming that the $D$ meson is produced purely by the $\bquark\to\cquark$ transition. 
A random amount of background $N^{\rm(bck)}_{\alpha (i)}$ is generated according to a Poisson distribution with mean $\langle N^{\rm(bck)}_{\alpha (i)}\rangle$ and is added to the signal yield. 
The expected background yields are then accounted for in the Poisson terms for each bin entering the likelihood function used in the fit to determine $\gamma$. 

\begin{figure}[htb]
  \centering
  \includegraphics[width=0.45\textwidth]{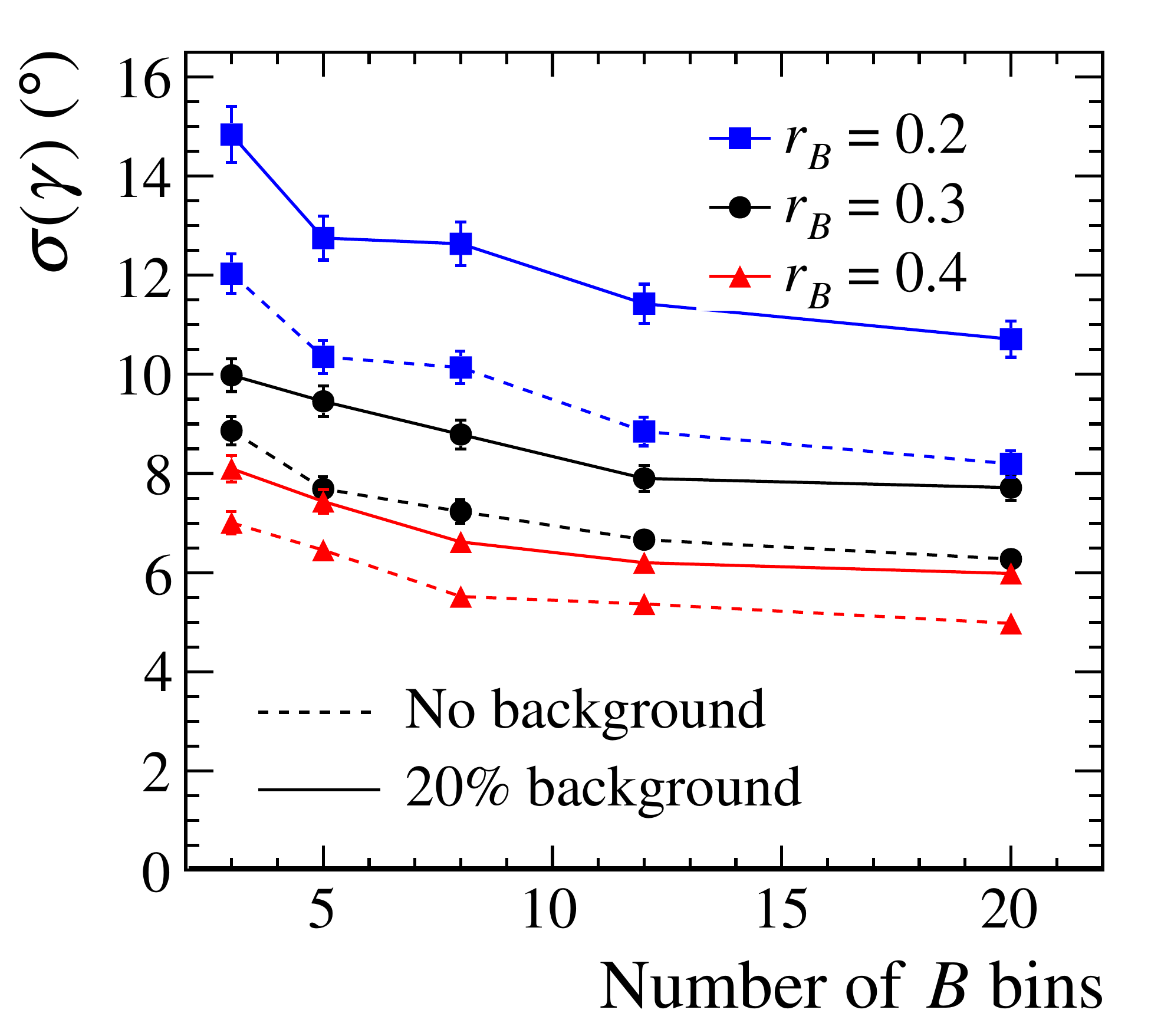}
  \put(-35, 45){(a)}
  \hspace{0.05\textwidth}
  \includegraphics[width=0.45\textwidth]{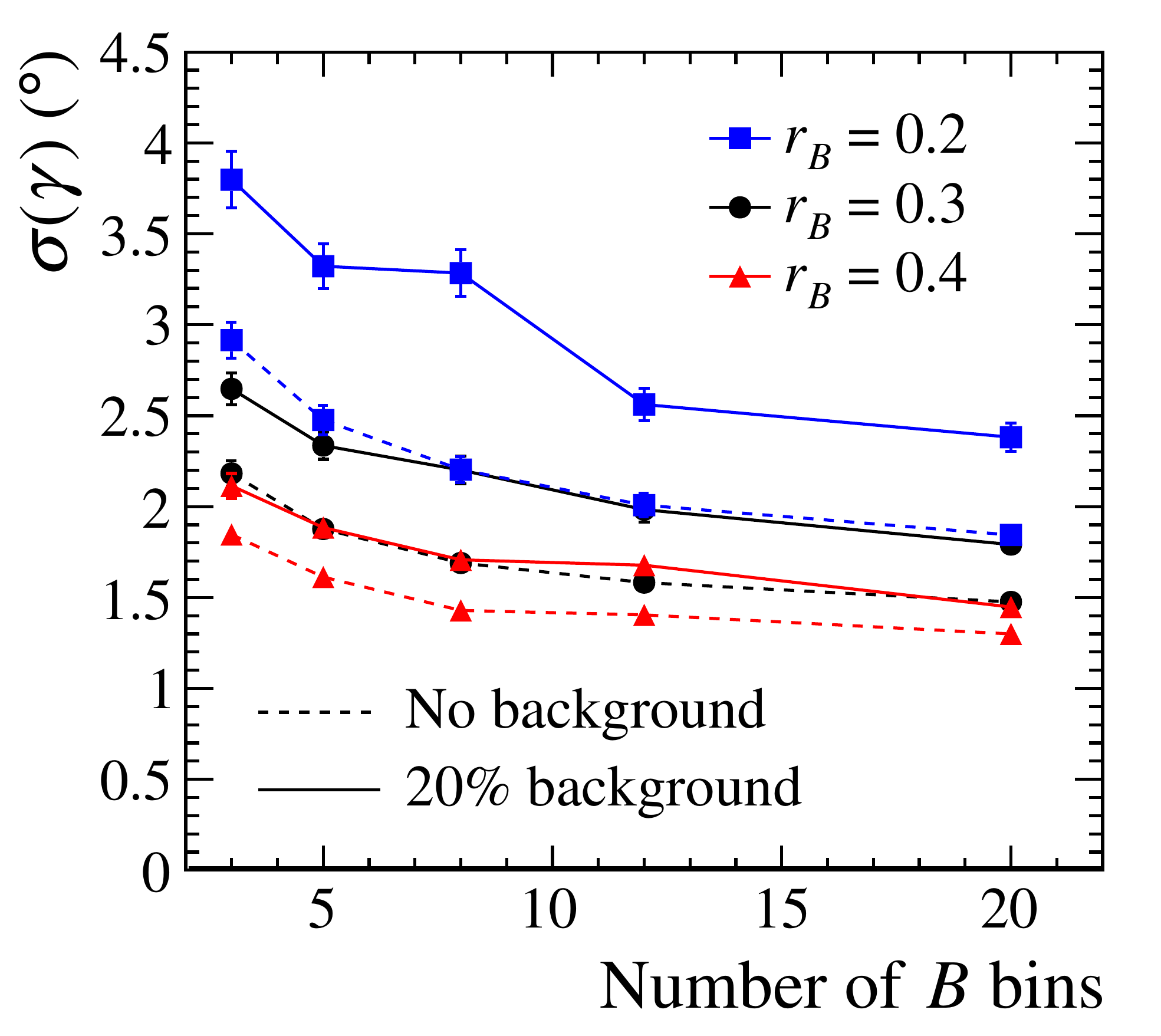}
  \put(-35, 45){(b)}

  \caption{
    Sensitivity to $\gamma$ in the (a) ``Run~I+II'' and (b) ``50\invfb'' scenarios, as a function of number of bins 
    of the $B$ Dalitz plot, shown for $r_B = 0.2$ (blue), $0.3$ (black) and $0.4$ (red), 
    without background (dashed line) and with the expected amount of $\Bs \to \Dstar \Km\pip$ background (solid line).
    The lines joining the points are added simply to guide the eye.
    Note that the points with no background correspond to those in Fig.~\ref{fig:rB-sensitivity} without the $D \to \Km\pip$ channel.
  }
  \label{fig:Bs-background-effect}
\end{figure}

The comparison of the sensitivity to $\gamma$ with and without the $\Bs \to \Dstar \Km\pip$ background included is shown in Fig.~\ref{fig:Bs-background-effect}. 
A deterioration in precision is seen in all scenarios, although the effect is larger for smaller values of $r_B$.
The size of the effect, around $10\%$, is significant but not large enough to threaten the viability of the method.
It may be possible to ameliorate the impact in an experimental analysis through selection requirements that discriminate against $\Bs \to \Dstar \Km\pip$ background or by taking the presence of the background into account in the determination of the binning scheme~\cite{Bondar:2008hh,Libby:2010nu}.

\subsection{\boldmath Impact of mismodelling of the $B$ decay amplitudes}

While the optimal binning of the $B$ decay phase space depends on the model, the measurement is unbiased even if the model used to define the binning differs from the true amplitude.
However, the statistical uncertainty of the measurement might be affected by mismodelling.
This effect is investigated by using alternative models for the binning optimisation with Eq.~(\ref{eq:binning-quality-simple}), while the pseudoexperiments are always generated according to the baseline model. 

\begin{figure}[htb]
  \centering
  \includegraphics[width=0.45\textwidth]{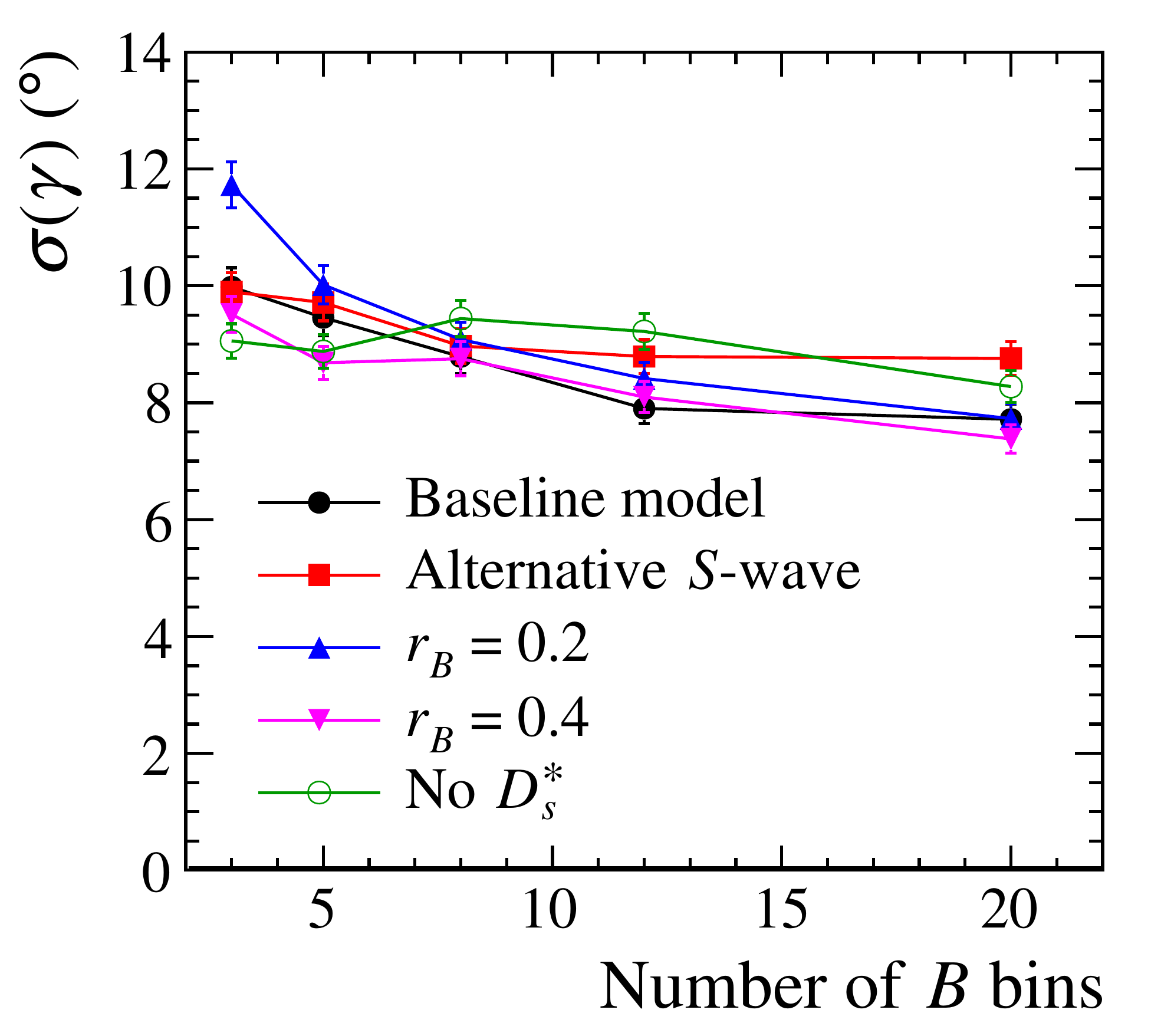}
  \put(-35, 45){(a)}
  \hspace{0.05\textwidth}
  \includegraphics[width=0.45\textwidth]{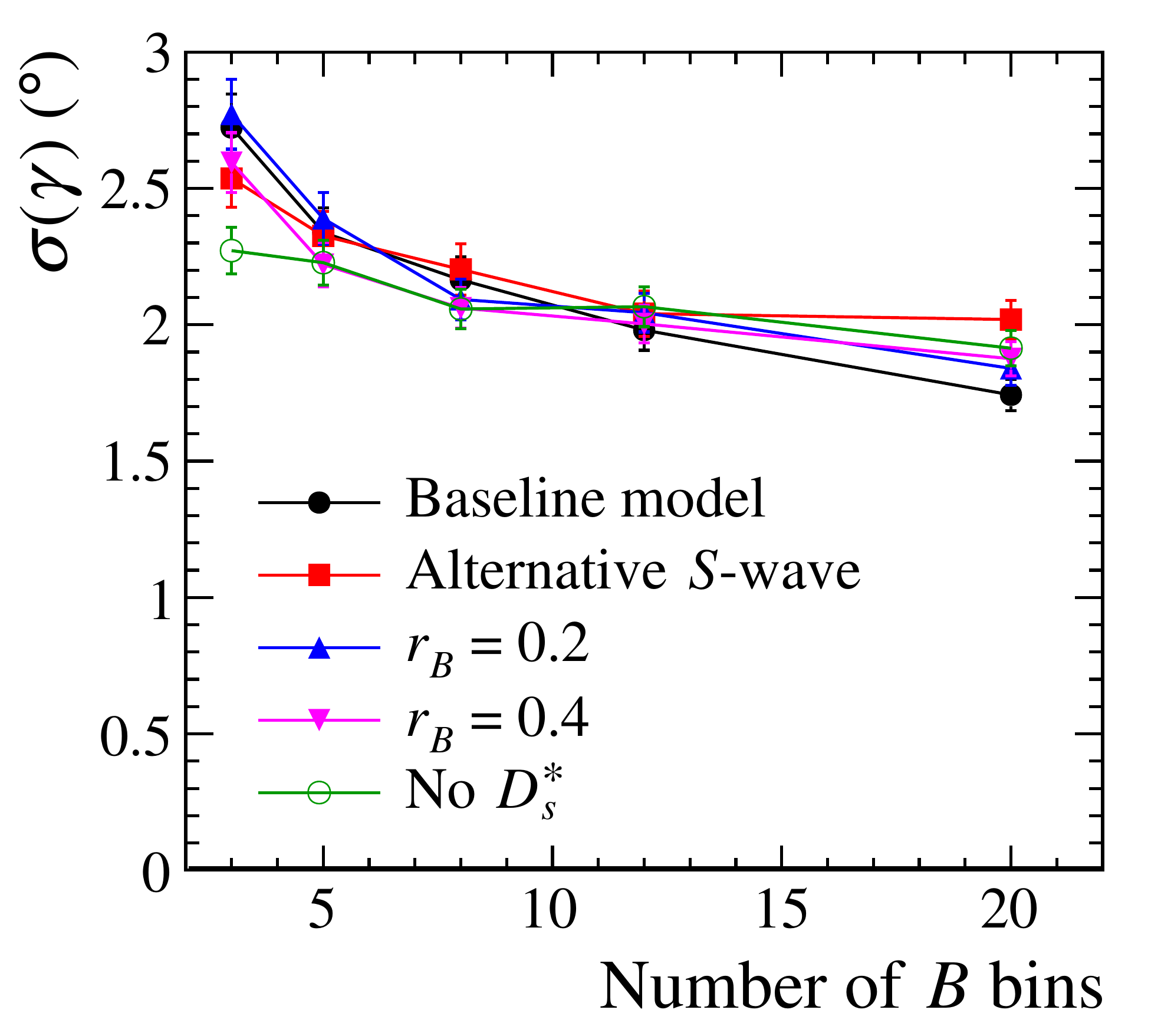}
  \put(-35, 45){(b)}

  \caption{
    Sensitivity to $\gamma$ in the (a) ``Run~I+II'' and (b) ``50\invfb'' scenarios, as a function of number of 
    bins of the $B$ Dalitz plot, with different $\Bz$ decay models used for binning optimisation. 
    The lines joining the points are added simply to guide the eye.
  }
  \label{fig:model_variations}
\end{figure}

It is expected that most aspects of the favoured $b \to c$ amplitude will be well known from the favoured mode, therefore most of the model variations considered relate to the suppressed amplitude only.
These include removing the $D_s^*(2700)$ state, as well as using $r_B=0.2$ or $0.4$ instead of the baseline $r_B=0.3$. 
There is also uncertainty related to the modelling of the broad $K\pi$ and $D\pi$ $S$-wave components (for the former appearing in both favoured and suppressed amplitudes; for the latter only in the favoured mode).
The impact of using alternative $S$-wave lineshapes in the binning optimisation to those in the generation is therefore considered, in a similar way to Ref.~\cite{LHCb-PAPER-2015-059}.
The results of this study are shown in Fig.~\ref{fig:model_variations}.
Event samples are generated with the baseline model and with $\Bs\to D^*\Km\pip$ background included at the expected level. 
The possible impact on the sensitivity to $\gamma$ is at the level of $10\%$, which is considered sufficiently small not to be a major concern.

\section{Summary}
\label{sec:summary}

The model-independent double Dalitz plot analysis of $\Bz\to D\Km\pip$ with, at least, $D\to \KS\pip\pim$ decays provides an attractive approach to the measurement of the angle $\gamma$ of the CKM Unitarity Triangle.
Using recently published information on the favoured and suppressed $B$ decay amplitudes~\cite{LHCb-PAPER-2015-017,LHCb-PAPER-2015-059}, the potential sensitivity of the method has been examined.
It is seen that sensitivities of around $8^\circ$ and $2^\circ$ can be expected for LHCb data samples corresponding to the expected amount of data collected at the end of the LHC Run~II and after $50\invfb$ have been collected.
These values are only around a factor of two larger than those expected from the combination of many results from LHCb~\cite{LHCb-PAPER-2012-031}, demonstrating that this method can have a significant impact.
The sensitivity depends strongly on the ratio of magnitudes of suppressed and favoured $B$ decay amplitudes, which is not yet well-known.
The dependence on the choice of model for the binning and the impact of background have been shown to be modest.
Thus, the major sources of systematic uncertainty that affect the determination of $\gamma$ from amplitude analysis of $\Bz \to D\Kp\pim$ decays~\cite{LHCb-PAPER-2015-059} are much less significant in the model-independent double Dalitz plot approach.
The method does not depend strongly on external constraints on the hadronic parameters $c_i$ and $s_i$ associated with the $D\to \KS\pip\pim$ decay, in contrast to the model-independent analysis for $\Bp\to D\Kp$ with multibody $D$ decays.
The double Dalitz plot approach is expected also to be relevant for the Belle~II experiment, where there will be no background from $\Bs \to \Dstar \Km\pip$ decays.
Further improvement in sensitivity may be achieved by optimising the binning taking backgrounds into account, or by adding further $D$ decay modes to the analysis, using the formalism set out in this paper.

\section*{Acknowledgements}

The authors wish to thank their colleagues on the LHCb experiment for the fruitful and enjoyable collaboration that inspired this study.
In particular, they would like to thank Matt Kenzie, Dan Johnson and Mark Whitehead for helpful comments on the manuscript.
This work is supported by the Science and Technology Facilities Council and by the English-Speaking Union (United Kingdom).

\addcontentsline{toc}{section}{References}
\setboolean{inbibliography}{true}
\bibliographystyle{LHCb}
\bibliography{references,main,LHCb-PAPER,LHCb-CONF,LHCb-DP,LHCb-TDR}

\end{document}